\documentclass{article}[12pt]
\usepackage{amsmath, amssymb,amsfonts}

 \newtheorem{thm}{Theorem}[section]

 \newtheorem{defn}[thm]{Definition}
 \numberwithin{equation}{section}

\newcommand{\lip}{\mathop{\mathrm{l.i.p}}}
\newcommand{\rot}{\mathrm{rot}\,}

\begin{document}

\title{The research program of Stochastic Deformation (with a view toward Geometric Mechanics)}

\author{Jean-Claude Zambrini\\
Grupo de F\'{i}sica-Matem\'atica da Universidade de Lisboa\\
Av. Prof. Gama Pinto, 2, 1649-003 Lisboa, Portugal\\
e-mail: zambrini@cii.fc.ul.pt }

\date{}

\maketitle

\begin{abstract}
We give an overview of a program of Stochastic Deformation of Classical Mechanics and the Calculus of Variations, strongly inspired by the quantization method.

\end{abstract}

\section{Introduction}

The program of Stochastic Deformation was born in 1984-5 as an attempt to understand the paradoxical probabilistic structures involved in quantum mechanics \cite{Zamb86}. In the work it was shown that although it is possible to directly associate diffusion processes with some elementary quantum systems (particles in potentials), their properties are irreducibly different from the predictions of quantum theory. For instance, their multi-time correlations are without any mathematical relations with the quantum ones. In addition the diffusions do not carry any quantum symmetries. The origin of these problems lies in the fact that the diffusions are not only driven by the given potentials but also by additional nonlocal ones, hidden within their original construction. No probabilistically consistent and physically relevant direct approach to quantum theory is therefore available today. Still, the method introduced in \cite{Zamb86} to construct this type of diffusions was quite unusual providing solutions of a new kind of stochastic boundary value problem, much more general than its original aim. So, on the positive side, a fresh mathematical reinterpretation of Feynman's path integral approach to the quantization problem was indeed accessible, providing a general method to produce quantum-like probability measures with qualitative properties quite distinct from what we are used to in statistical physics.

It is our intention to describe here the main features of this method. No familiarity with Feynman's path integral approach or even quantum mechanics itself, is required. We shall summarize the basic elements, allowing the reader to understand why our construction is a rigorous version of this approach.

Quantum Theory provides for a kind of deformation of Classical Mechanics, but unfortunately not a probabilistic one. One way to see our construction is, precisely, as such a quantum-like stochastic deformation. In this respect, it should also be of interest in stochastic approaches to Geometric Mechanics \cite{LazOrt88}.

It may also be worth recalling that the project to make sense of the Feynman Path Integral method (and not simply of a few of his formulas) is still a largely open problem with a potentially devastating generality for probabilists. Who can doubt, indeed, after all these years and so many applications far beyond what Feynman could have imagined in the fifties, that a mathematically consistent version of this method should exist ?

This overview of Stochastic Deformation is organized as follows:

Section 2 summarizes the original ideas of Feynman's reinterpretation of elementary quantum mechanics. For a probabilist, they look like a (very) informal version of Stochastic Analysis, with a twist regarding boundary conditions.This twist will prove to be fundamental for their mathematical interpretation.

Section 3 provides a probabilistic counterpart of Feynman's approach, i.e. a kind of stochastic boundary value problem whose conditions of existence and uniqueness of solutions are specified.

In Section 4 the random dynamics of the relevant class of processes, together with its associated symmetries, will be described in bothLagrangian and Hamiltonian form.

The last Section is devoted to some computational and geometric aspects of our Stochastic Deformation.

\section{Overview of Feynman Path Integral approach to quantum theory}

We consider the example of a system of a single unit mass (charged) particle in a (bounded below) scalar potential $V(q)$ and a smooth vector potential $a(q)$. According to elementary quantum mechanics \cite{Grif04}, the Hamiltonian observable of such a system is a densely defined self-adjoint operator (in $L^2(\mathbb{R}^3)$ for simplicity)
\begin{equation}\label{eq2.1}
H=\frac{1}{2}(P-a(Q))^2+V(Q)
\end{equation}
where the position and momentum observables are defined respectively on appropriate dense domains as
\begin{equation}\label{eq2.2}
\begin{split}
Q:\mathcal{D}_Q\subset L^2&\to L^2\\
\psi(q)&\mapsto q\psi(q)
\end{split}
\end{equation}
\begin{equation}\label{eq2.3}
\begin{split}
P:\mathcal{D}_p\subset L^2&\to L^2\\
\psi(q)&\mapsto -i\hbar\nabla\psi(q)
\end{split}
\end{equation}
for $\hbar$ a positive constant.

The initial state $\psi$ of the system evolves in an unitary way, $U_t\psi=\psi_t$, $U_t=e^{-\frac{i}{\hbar} t H}$, so that $\psi_t$ solves 
\begin{equation}\label{eq2.4}
i\hbar\frac{d}{dt}\psi_t=H\psi_t
\end{equation}

Any observable, in fact, is a densely defined self-adjoint operator O. The contact with (the language, if not the mathematics, of) probability theory is made by the definition of the ``mean value of O in the state $\psi_t$" at time $t$:
\begin{equation}\label{eq2.5}
\langle O \rangle_{\psi_t}\equiv\langle \psi_t | O\psi_t\rangle=\langle\psi | O_t \psi\rangle\equiv \langle O_t\rangle_\psi,\quad O_t=U^{+}_{t} O U_t
\end{equation}
where $\langle\cdot | \cdot \rangle$ denotes the $L^2$ scalar product, antilinear on the left (with associated norm $\|\cdot\|$), and $+$ is the adjoint.

Eq \eqref{eq2.5} is a dual expression of the dynamics of an observable. According to Dirac's ``correspondence principle", to the Poisson bracket of two classical observables $b$, $c$ (i.e. continuous functions of $(q,p)\in T^\ast\mathbb{R}^3$) should correspond the commutator of their quantum counterparts $B$ and $C$:
\begin{equation*}
\{b,c\}\longleftrightarrow\frac{1}{i\hbar}[B,C].
\end{equation*}
Here we shall not need to worry about the fact that such a correspondence cannot be raised to the status of homomorphism of Lie algebra. It will be sufficient to observe that, indeed, to the dynamical law of the (possibly time dependent) classical observable $b$,
\begin{equation*}
\frac{db}{dt}=\frac{\partial b}{\partial t}+\{b,h\}
\end{equation*}
where $h$ denotes the classical Hamiltonian of the system corresponds Heisenberg's equation:
\begin{equation*}
\frac{dB_t}{dt}=\frac{\partial B_t}{\partial t}+\frac{1}{i\hbar}[B_t,H].
\end{equation*}
Applying this to $B_t=Q_t$ then $P_t$, for $H$ as in Eq \eqref{eq2.1} and taking the mean value in the state $\psi$ we obtain
\begin{equation}\label{eq2.6}
\begin{cases}
\frac{d}{dt}\langle Q_t\rangle_\psi=\langle P_t - a_t\rangle_\psi\\
\frac{d}{dt}\langle P_t - a\rangle_\psi=\langle (P_t - a_t)\wedge \mathrm{rot} \,a_t - \nabla V (Q_t)\rangle_\psi
\end{cases}
\end{equation}

But of what kind of random variables, really, those expressions are the mean values? We do not know; in fact no probability space has ever been defined in the first place. The only hint at probability theory, in quantum mechanics is Von Neumann's Axiom of ``Quantum Static" according to which if one performs a measurement of an observable O for a system in state $\psi$, the absolute probability to find a result $\leqslant \lambda\in\mathbb{R}$ is 
\begin{equation}\label{eq2.7}
\langle E^O(\lambda)\rangle_\psi=\|E^O(\lambda)\psi\|^2
\end{equation}
where $E^O(\lambda)$ denotes the spectral family of orthogonal projections of $O$. For instance, when $O=Q$, and any $\lambda$ in the interval $[a,b]$,
\begin{equation}\label{eq2.8}
\|E^Q([a,b])\psi\|^2=\int_{[a,b]} |\psi(q)|^2 dq
\end{equation}
called Born interpretation of the wave function $\psi$. In this sense Eqs (\ref{eq2.5}--\ref{eq2.6}) are understandable, if not justified probabilistically. Even for the simplest quantum systems (Eq \eqref{eq2.1} with $a=V=0$, i.e the ``free case") there is no underlying concept of space-time trajectory. The justification of this prohibition lies in the uncertainty relation.

For $\psi$ in the domains of $Q$ and $P$,
\begin{equation}\label{eq2.9}
(Q_j P_k-P_k Q_j)\psi=i\hbar\delta_{jk}, \quad 1\leqslant j,k\leqslant 3
\end{equation}
is interpreted as the impossibility to localize experimentally the position and the momentum simultaneously, i.e to define a trajectory as in classical Hamiltonian mechanics.

Feynman transformed qualitatively these shaky relations between Quantum Physics and Probability Theory \cite{Feyn65}.

Let us consider a classical (conservative) system of Lagrangian  $L$, in dynamical evolution on the time interval $I=[s,u]$. If its configuration at time $t\in I$ is denoted by $\omega(t)$, the Action functional of this system is defined by 
\begin{equation}\label{eq2.10}
S_L[\omega(\cdot); u-s]=\int^u_s L(\omega(t), \dot{\omega}(t)) dt
\end{equation}

For our system whose quantum Hamiltonian is \eqref{eq2.1},
\begin{equation}\label{eq2.11}
L(\omega,\dot{\omega})=\frac{1}{2}|\dot{\omega}|^2-V(\omega)+a(\omega)\cdot\dot{\omega}.
\end{equation}
Feynman starts from two states $\varphi$, $\psi$, say at the final time $u$ of $I$ and rewrites $\langle\varphi_u | \psi_u\rangle\in\mathbb{C}$ in terms of the unitary evolution kernel applied to $\psi_s$:
\begin{equation}\label{eq2.12}
\int\int\psi_s(x)(e^{-\frac{i}{\hbar}(u-s)H})(x,z) \overline{\varphi}_u(z)\, dx\,dz
\end{equation}
where $\overline{\varphi}$ denotes the complex conjugate.

The key point is that he reinterprets this as the following ``Path Integral":
\begin{equation}\label{eq2.13}
\int\int_{\Omega^{z,u}_{x,s}}\psi_s(x) e^{\frac{i}{\hbar}S_L[\omega(\cdot); u-s]}\mathcal{D}\omega\,\overline{\varphi}_u(z)\,dx\,dz
\end{equation}
In this symbolic expression, $\Omega^{z,u}_{x,s}$ means the set of continuous paths $\{\omega\in C([s,u],\mathbb{R}^3)|\omega(s)=x, \omega(u)=z\}$ and $\mathcal{D}\omega$ plays the role of a ``flat measure" $\prod_{s\leqslant t\leqslant u} d\omega(t)$ on this path space.

Consider any time $s<t<u$. Clearly $\psi_s$ can be regarded as initial condition of
\begin{equation}\label{eq2.14}
i\hbar\frac{\partial\psi}{\partial t}=H\psi,
\end{equation}
and $\overline{\varphi}_u$ as final boundary condition of the adjoint problem 
\begin{equation}\label{eq2.15}
-i\hbar\frac{\partial\overline{\varphi}}{\partial t}=H\overline{\varphi}
\end{equation}

Feynman calls ``transition amplitude" the (complex) expression \eqref{eq2.12} or \eqref{eq2.13}. He will use it as a kind of expectation with weight $\exp\frac{i}{\hbar} S_L$, and denotes it by $\langle\varphi|1|\psi\rangle$ or $\langle 1 \rangle_{S_L}$. It should be stressed that, in Feynman's view, \eqref{eq2.13} should simply be regarded as a short notation for a sum over time-discretized paths $\omega(t_j)=x_j$, $t_j=s+j\frac{(u-s)}{N}$, $1\leqslant j\leqslant N\in\mathbb{N}$. Along the same line, for ``any" functional $F[\omega(\cdot)]$ of such paths, the ``expectation" of $F$ is defined by
\begin{equation}\label{eq2.16}
\langle{F}\rangle_{S_L}=\int\int\int\psi_s(x)\; e^{\frac{i}{\hbar}S_L[\omega(\cdot); u-s]} F[\omega(\cdot)]\overline{\varphi}_u(z)\,\mathcal{D}\omega\,dx\,dz.
\end{equation}
This definition is the starting point of Feynman's ``Functional Calculus", whose main result is an Integration by parts formula \cite{Feyn65}:
\begin{equation}\label{eq2.17}
\langle{\delta\,F[\omega](\delta\omega)}\rangle_{S_L}=-\frac{i}{\hbar} \langle{F\delta S_L[\omega](\delta\omega)}\rangle_{S_L}
\end{equation}
where $\delta G[\omega](\delta\omega)$ denotes the Gateaux derivative of a functional $G$ on $\omega$ in the direction $\delta\omega$. Shaky as it is, mathematically, this formula deserves some interest since it is the ancestor of all the integration by parts formulas designed in Stochastic Analysis during the last 25 years ! 

For $F=1$, $L$ as in \eqref{eq2.11}, Feynman finds the path integral counterpart of \eqref{eq2.6}:
\begin{equation}\label{eq2.18}
\langle\ddot{\omega}\rangle_{S_L}=\langle\dot{\omega}\wedge\mathrm{rot}\, a(\omega)-\nabla V(\omega)\rangle_{S_L}
\end{equation}
where, regarding the meaning of the l.h.s, he observed that ``in the few examples with which we had experience, the substitution $\ddot{\omega}=\frac{1}{(\Delta t)^2}(\omega(t+\Delta t)-2\omega(t)+\omega(t-\Delta t))$ has been adequate".

The most revealing application of the relation \eqref{eq2.17}, however, is the one for $F[\omega]=\omega$ and some specific direction $\delta \omega$. Indeed, \eqref{eq2.17} reduces, then, to the (discretized) expression:
\begin{equation}\label{eq2.19}
\Bigg\langle{\omega_j(t)\left(\frac{\omega(t)-\omega(t-\Delta t)}{\Delta t}\right)_k}\Bigg\rangle_{S_L}-\Bigg\langle\left(\frac{\omega(t+\Delta t)-\omega(t)}{\Delta t}\right)_k \omega_j(t)\Bigg\rangle_{S_L}=i\hbar\delta_{jk}.
\end{equation}
This maybe the most fundamental discovery of Feynman approach, \cite{Feyn65} although it is almost never cited, for reasons to be understood afterwards. First, let us observe that \eqref{eq2.19} is a kinematical claim regarding the nature of trajectories. No dynamics is involved here. By construction, $\omega(\cdot)\in\Omega^{z,u}_{x,s}$, but what \eqref{eq2.19} suggests (since Feynman was aware that the desirable $\lim_{\Delta t\downarrow 0}$ is, to say the least, problematic) is that $\dot{\omega}(t_{-})$ should be different from $\dot{\omega}(t^{+})$, for any $t\in]s,u[$. In other words, the $\omega(\cdot)$ are indeed continuous quantum trajectories, but they are not differentiable.

Now consider $a=0$ in the Lagrangian \eqref{eq2.11}. Then, by definition of the classical momentum, $p=\frac{\partial L}{\partial\dot{\omega}}=\dot{\omega}$ if the configuration $q=\omega$. So Feynman regards \eqref{eq2.19} as the ``space-time" version of the uncertainty relation \eqref{eq2.9}, providing a deeply different interpretation from the regular one of Quantum Theory in Hilbert space: no reference to any limitation on experimental measurement is involved here.

Notice that, in this perspective and for Lagrangians as before, it would make perfect sense to interpret Quantum Mechanics as a Stochastic Deformation of Classical Mechanics for smooth paths. Clearly, $\hbar$ is the deformation parameter. However, even $S_L[\omega(\cdot)]$ would become singular along such ``quantum paths" since $\dot{\omega}$ does not make sense anymore. 

Let us come back, for instance, on our question, after Eq \eqref{eq2.6}: of what kind of random variables are those equations mean values? Take the above elementary momentum, for instance. By \eqref{eq2.6} (with $a=0$) it should satisfy, in particular, after quantization,
\begin{equation}\label{eq2.20}
\frac{d}{dt}\langle{Q_t}\rangle_{\psi}=\langle{P_t}\rangle_{\psi}.
\end{equation}
According to Feynman, however, this corresponds to 
\begin{equation}\label{eq2.21}
\frac{d}{dt}\langle\omega(t)\rangle_{S_L}=``\lim_{\Delta t\downarrow 0}"\Bigg\langle\frac{\omega(t+\Delta t)-\omega(t)}{\Delta t}\Bigg\rangle_{S_L}
\end{equation}
where $\omega(\cdot)$ denotes some, yet unspecified, random process.

The right hand side can be computed as a difference of 2 functionals like \eqref{eq2.16}, using Eqs (\ref{eq2.14}--\ref{eq2.15}). On the other hand, of course, according to regular quantum theory, and using \eqref{eq2.3}, we know that 
\begin{equation}\label{eq2.22}
\langle{P_t}\rangle_{\psi}=\langle{P}\rangle_{\psi_t}=\int\overline{\psi}_t(-i\hbar\nabla\psi_t)dq=\int\overline{\psi}_t\psi_t\left(-i\hbar\frac{\nabla\psi_t}{\psi_t}\right)\,dq.
\end{equation}
In this way Feynman reinterprets each quantum observable as a specific function of his ``random process" $\omega(\cdot)$, here $-i\hbar\nabla\log\psi_t$ for $P$. Unfortunately, R.H. Cameron proved in 1960 (\cite{Cam60}) that $e^{\frac{i}{\hbar}S_L}\mathcal{D}\omega$ does not make any sense as a countably additive measure on the path space underlying \eqref{eq2.13}. This means that there is no such thing as the ``expectation" $\langle{\;}\rangle_{S_L}$ of \eqref{eq2.16}, and no $\lim_{\Delta t\downarrow 0}$ in expressions like \eqref{eq2.19} or \eqref{eq2.21}.

A complex measure similar to the Wiener one but with a purely complex variance, like the one used informally by Feynman would be in particular of infinite total variation and therefore not appropriate for his quantum purpose.

The challenge we face is therefore to preserve the essential of Feynman's approach but with well defined probability measures on path spaces. In order to tackle this, we will first summarize some of the key qualitative aspects to preserve.

\begin{enumerate}
\item Assume that there is a (filtered) probability space $(\Omega, \sigma,P)$ where such a stochastic process, $\omega(t)=X(t)$, $t\in I$, is well defined.

Then the uncertainty relation \eqref{eq2.19} should mean 
\begin{multline}\label{eq2.23}
E\left\{X_j(t)\lim_{\Delta t\downarrow 0} E_t\left[\frac{X(t)-X(t-\Delta t)}{\Delta t}\right]_k\right\}-\\
E\left\{\lim_{\Delta t\downarrow 0} E_t\left[\frac{X(t+\Delta t)-X(t)}{\Delta t}\right]_k X_j(t)\right\}=\hbar\delta_{jk}
\end{multline}
where $E_t$ should be a conditional expectation since we know (cf. \eqref{eq2.22}) that the momentum has to become a function of the process $X(t)$. This means, in particular, that $X(t)$ should be Markovian. Several remarks may help here. First, even for the simplest process we could associate with the simplest Hamiltonian \eqref{eq2.1} (the ``free" one, $H_0$, where $a=V=0$) namely a Wiener $X(t)=W(t)$, and even using the weakest type of convergence of random variables, the convergence in probability, the ``forward" derivative $\lim_{\Delta t\downarrow 0}\left(\frac{W(t+\Delta t)-W(t)}{\Delta t}\right)$ involved in \eqref{eq2.19} does not exist , i.e. the Wiener trajectories are, indeed, not differentiable at any time $t$. However, the limit of conditional expectations of such derivatives (playing the role of Feynman's $<\cdot>_{SL}$), denoted afterwards by $D_t X$, should be well defined random variables. We shall denote by $D^\ast_t X$ the conditional expectation  of the backward derivative needed as well in \eqref{eq2.23}. Note that the backward increment, given $X(t)$, involved in $D^\ast_t X$ is not available, in general, if we are only given the past information until time $t$, modelized by an increasing sigma algebra $\mathcal{P}_t$. To make sense of Eq \eqref{eq2.23}, both limits should, of course, be distinct. Following R.H. Cameron, we cannot hope to produce complex measures satisfying Eq \eqref{eq2.19}. We would content ourselves with real measures. 

At this point a brief recollection of the fundamental notion of conditional expectation, a kind of partial averaging, may also be useful. For a given probability space $(\Omega, \sigma, P)$ the conditional expectation of a random variable $\xi$ with respect to a sigma-algebra $\mathcal{P}\subset \sigma$ is another random variable, $E[\xi|\mathcal{P}]$, defined by the property $E[E[\xi |\mathcal{P}]]=E[\xi]$, for $E$ the absolute expectation.

In this survey, given a process $X(t)$, we denote by $E_t[\cdot]$ the conditional expectation $E[\cdot | X(t)]$. When the condition is $X(t)=x$ then $E[\cdot | X(t)=x]$ becomes a (non random) function of $x$ and possibly $t$, denoted by $E_{xt}[\cdot]$.

\item We need to preserve a probabilistic counterpart of the time symmetry in quantum physics (at least for systems whose Lagrangian is time independent) involved in Born interpretation \eqref{eq2.8}. Indeed, the relation between $\psi$ and $\overline{\psi}$ can be regarded as a time reversal (cf. \eqref{eq2.14}--\eqref{eq2.15}) therefore their product, in Born interpretation, is unchanged under this symmetry.

\item Although  we shall treat here exclusively the class of elementary Lagrangian systems \eqref{eq2.11} considered by Feynman, our construction should rely on general principles compatible with any physical system.
\end{enumerate}

\vspace{0.2cm}
Before describing the program of Stochastic Deformation, a few words about the internal evolution of Stochastic Analysis itself.

After the pioneering works of Wiener and It\^o, this field made great progress since 1980 but there is one aspect where it did not; with the notable exception of some approaches of Stochastic Control (we shall come back on this), the field suffered from a chronic lack of dynamical content (in the classical sense of dynamical systems theory). In It\^o's original perspective stochastic differential equations are stochastic deformation of ordinary differential equations, still the comparison of the history of the two fields is revealing. For (second order) ODE a single, very hard, dynamical problem became the motor of all scientific progress: the $N$-body problem. Nothing like it was ever considered in Stochastic Analysis. This explains why very basic notions of ODE's theory, like the one of integrability, for instance, are lacking in Stochastic Analysis.

Here is a (dynamical) ``paradox" mentioned by Krzysztof Burdzy. Let $\phi:[0,T]\to\mathbb{R}$ such that $\sup_{t\in[0,T]} |\ddot{\phi}(t)|<\infty$. If $W_t$ denotes the Brownian motion, it is known that the probability 
\begin{multline*}
P\{\phi(t)-\varepsilon< W_t<\phi(t)+\varepsilon,\; \forall t\in[0,T]\}\sim\\
 c(\varepsilon)\exp-\frac{1}{2}\int^T_0 (\dot{\phi}(\tau))^2 d\tau\equiv F[\phi(\cdot)]
\end{multline*}
The functional $F[\phi]$ is maximized by $\phi(\tau)=0$\; $\forall \tau\in[0,T]$ i.e, in particular, the solution of the second order ODE $\ddot{\phi}(t)=0$ with $\phi(0)=\dot{\phi}(0)=0$. Burdzy compares this most likely shape of a Brownian path to the erratic Microsoft Stock price during 5  years. What part of Brownian dynamics is really captured by the solution of such a classical variational principle? After all, the roughness of Brownian paths is already incompatible with any notion of classical kinematics. In fact we shall need precisely to generalize those classical notions of kinematics and dynamics before offering an answer to this paradox. Notice that, according to Feynman, this is the free case $V=a=0$ in \eqref{eq2.11} and the dynamical equation of this system should indeed be (cf \eqref{eq2.18}) $\langle\ddot{\omega}\rangle_{S_L}=0$, whatever meaning can be given to $\langle\cdot\rangle_{S_L}$.

\section{Probabilistic counterpart of Feynman's approach}
\setcounter{equation}{0}

To make sense of \eqref{eq2.23}, in the form 
\begin{multline}\label{eq3.1}
E\bigg[X_j(t)\lim_{\Delta t\downarrow 0} E_t\left[\frac{X(t)-X(t-\Delta t)}{\Delta t}\right]_k-\\\lim_{\Delta t\downarrow 0} E_t\left[\frac{X(t+\Delta t)-X(t)}{\Delta t}\right]_k X_j(t)\bigg]=\hbar\delta_{jk}
\end{multline}
we need two filtrations, to take into account not only the usual past information on a time interval $I\supset[s,u]$, i.e an increasing one $\mathcal{P}_t, t\in I$, but also a decreasing filtration $\mathcal{F}_t$ taking into account the future. The underlying filtered probability space should, therefore, be of the form $(\Omega,\sigma, \{\mathcal{P}_t\}, \{\mathcal{F}_t\}, P)$ with $t\in[s,u]$.

Feynman's time discretized interpretation of the left hand side of his dynamical Eq \eqref{eq2.18} suggests to limit ourselves to processes $X_t$ such that, for any bounded measurable $f$ and any $s\leqslant s_1<t<t_1\leqslant u$,  
\begin{equation}\label{eq3.2}
E\left[f(X_t)|\mathcal{P}_{s_1}\cup\mathcal{F}_{t_1}\right]=E\left[f(X_t)|X(s_1), X(t_1)\right].
\end{equation}
This property is what we call now Local Markov (or two-sided Markov). But it was introduced in 1932 by Sergei Bernstein, who named it ``reciprocal" \cite{Bern32}. His motivation was a remark of E. Schr\"odinger, a year earlier, inspired by the foundations of quantum mechanics, and which seems to be at the origin of all the notions of stochastic reversibility known today to probabilists \cite{Schr32}. 

To keep track, as suggested by Feynman, of the past and future informations about the system, the traditional Markovian transition probability should be replaced by a more symmetric measure $Q$, named after Bernstein:

$A\mapsto Q(s,x,t,A,u,z)$, $\forall x,z\in\mathbb{R}$, $s<t<u$ in $I$, measurable in $x,z$ with $A\in\mathcal{B}(\mathbb{R})$, where $\mathcal{B}$ is the Borel sigma-field.

For $Q$ there is a 3 points analogue of Chapman-Kolmogorov property, such that, for $X(u)=z$ fixed, $Q$ becomes a forward Markov transition and for $X(s)=x$ fixed $Q$ reduces to a backward Markov property (let us recall, with A.D. Wentzell \cite{Went81}, that these are just two of the 64 ways to express Markov property !).

Of course, without fixing the starting or ending point, $X(\cdot)$ will only be a Bernstein process, satisfying \eqref{eq3.2} and not a Markovian one. 

Let us denote by $M$ the joint probability measure on $\mathcal{B}\times\mathcal{B}$ for the pair of initial and final random variables.

Then B. Jamison (1974 \cite{Jam74}) proved the following (cf. \cite{Zamb86} or \cite{Cruz91}, for the symmetric version involving $\mathcal{P}_t$ and $\mathcal{F}_t$, appropriate to the relation with Feynman's approach to quantum mechanics).

\begin{thm}
For a given Bernstein transition $Q$ and a given joint measure $M$,
\begin{description}
\item{a)} There is a unique probability measure $P_M$ such that under $P_M$, $X(t)$, $t\in[s,u]$, satisfies Bernstein property \eqref{eq3.2}.
\item{b)} $P_M(X(s)\in A_s, X(u)\in A_u)=M(A_s\times A_u)$ for any Borelians in $\mathcal{B}$, the Borel tribe of $\mathbb{R}^3$.
\item{c)} $P_M(X(s)\in A_s, X(t_1)\in A_1,\dots,X(t_n)\in A_n, X(u)\in A_u)$
\begin{multline*}
=\int_{A_s\times A_u} dM(x,z)\int_{A_1} Q(s,x,t_1,dx_1,u,z)\int_{A_2}\dots\\\int_{A_n} Q(t_{n-1},q_{n-1},t_n,dq_n,u,z)
\end{multline*}
for $s\leqslant t_1\leqslant t_2\leqslant\dots\leqslant t_n<u$ and $A_i\in\mathcal{B}$, $i=1,\dots,n$.
\end{description}
\end{thm}

The final random variable has been fixed here so that, as said before, $Q$ has the properties of a forward Markovian transition but c) would hold as well with a fixed initial $X(s)=x$. In other words, the construction is perfectly symmetric with respect to the past and future informations, as it should be.

Jamison also proved that only one joint probability measure $M=M_m$ ($M_m$ for Markov) turn $X(t)$ into a Markovian and not only a Bernstein process. Using the same notation as in $\S 1$ but, this time, for the (strongly continuous contraction) semigroup generated by the lower bounded operator $H$, $M_m$ is of the form 
\begin{align}\label{eq3.3}
M_m(A_s\times A_u)=\int_{A_s\times A_u}\eta^{\ast}_s (x)\left(e^{-\frac{1}{\hbar}(u-s)H}\right)(x,z)\eta_u(z)\,dx\,dz
\end{align}
where $\eta^\ast_s(x)$ and $\eta_u(z)$ are two positive (not necessarily bounded) functions to be determined later. This expression should be compared with Eq \eqref{eq2.12}. Now, with 
\begin{align*}
Q(s,x,t,dq,u,z)=h(s,x,u,z)^{-1}\cdot h(s,x,t,q)\cdot h(t,q,u,z) \cdot dq,\; s<t<u
\end{align*}
where the handy notation $(e^{-\frac{1}{\hbar}(u-s)H}) (x,z)=h(s,x,u,z)$ has been used, the substitution of \eqref{eq3.3} in the above finite dimensional distributions c) provides, after simplifications, the finite dimensional distributions:
\begin{align*}
\rho_n(dx_1,t_1,dx_2,t_2,\dots,dx_n,t_n),\; s<t_1<t_2<\dots<t_n<u
\end{align*}
\begin{equation}\label{eq3.4}
=\int_{A_s\times A_u}\eta^\ast_s(x) h(s,dx,t_1,dx_1)\dots h(t_n,dx_n,u,dz)\eta_u(z)\,dx\,dz
\end{equation}
Now define the following densities of the forward and backward transition probabilities 
\begin{align}
&P(t,x,u,dz)=h(t,x,u,z)\,\frac{\eta_u(z)}{\eta_t(x)}\, dz\quad \forall\,t\leqslant u\label{eq3.5}\\
\intertext{where}
&\eta_t(x)=\int h(t,x,u,z)\eta_u(z)\,dz,\label{eq3.6}
\intertext{and}
&P^\ast(s,dy,t,x)=\frac{\eta^\ast_s(y)}{\eta^\ast_t(x)} h^\ast(t,x,s,y)\,dy\qquad \forall\, t\geqslant s\label{eq3.7}\\
\intertext{where}
&\eta^\ast_t(x)=\int\eta^\ast_s (y)h^\ast(t,x,s,y)\,dy=\int\eta^\ast_s(y) h(s,y,t,x)\,dy\label{eq3.8}
\end{align}
and the classical relation $h^\ast(t,x,s,y)=h(s,y,t,x)$ between integral kernels of two adjoint parabolic equations (for $H$ not necessarily symmetric) has been used. Then it is easy to check that \eqref{eq3.4} coincides with the finite dimensional distributions of a forward Markovian process of initial probability density $\eta^\ast_s(x)\eta_s(x)$ and transition probability density \eqref{eq3.5} or, equivalently, of a backward Markovian with final probability density $\eta^\ast_u(z)\eta_u(z)$ and backward transition density of the form \eqref{eq3.7}.

As a matter of fact, $\forall\,t\in[s,u]$, it is true that 
\begin{equation}\label{eq3.9}
P(X(t)\in A)=\int_A \eta^\ast_t\eta_t(x)\,dx,\quad A\in\mathcal{B}
\end{equation}
where, as shown by \eqref{eq3.6} and \eqref{eq3.8}, $\eta^\ast_t$ and $\eta_t$ are two positive solutions of the two adjoint parabolic PDE, $s\leqslant t\leqslant u$
\begin{equation}\label{eq3.10}
\begin{cases}
-\hbar\frac{\partial\eta^\ast}{\partial t}=H^+\eta^\ast\\
\eta^\ast(s,x)=\eta^\ast_s(x)
\end{cases}
\end{equation}
and
\begin{equation}\label{eq3.11}
\begin{cases}
\hbar\frac{\partial\eta}{\partial t}=H\eta\\
\eta(u,x)=\eta_u(x).
\end{cases}
\end{equation}

Here, a comment is needed since we said that quantum Hamiltonians, like any observable, are self-adjoint and the first PDE of \eqref{eq3.10} involves the adjoint $H^+$ of $H$. Our example \eqref{eq2.1} illustrates this point. When written explicitly, Schr\"odinger equation \eqref{eq2.4} means
\begin{equation}\label{eq3.12}
i\hbar\frac{\partial\psi}{\partial t}=-\frac{\hbar^2}{2}\Delta\psi+\frac{i\hbar}{2}\nabla . a\psi+i\hbar a\nabla\psi+\frac{1}{2}|a|^2 \psi+V\psi
\end{equation}
The ``Euclidean version" of this corresponds to transform $t$ into $it$ and $a$ into $-iA$ so that the right hand side operator becomes, indeed, a non-symmetric operator
\begin{equation}\label{eq3.13}
H=-\frac{\hbar^2}{2}\Delta+\hbar A\nabla+\frac{\hbar}{2} \nabla . A-\frac{1}{2}|A|^2+V.
\end{equation}
Notice how close we are, in Eq \eqref{eq3.9}, to Born interpretation of the wave function $\psi$ in \eqref{eq2.8}. Informally, we have done $t\mapsto it$ and the above transformation of the vector potential $a$. The product structure of the probability density in \eqref{eq3.9} is fundamental; it expresses manifestly a kind of invariance under time reversal, more general than the one traditionally known by probabilists as ``reversibility". For instance, if $P_t(dx)$ denotes this probability at time $t$, it follows immediately from \eqref{eq3.5} and \eqref{eq3.7} that the following ``detailed balance" condition holds
\begin{equation}\label{eq3.14}
P_s(dx)P(s,x,u,dz)=P^\ast(s,dx,u,z) P_u(dz), s\leqslant u
\end{equation}
generalizing Kolmogorov's notion of reversibility \cite{Dob88} to non-stationary situations. In this paper, in fact, Kolmogorov refers to E. Schr\"odinger (1931,32) \cite{Schr32} who is at the very origin of our program of Stochastic Deformation. This has been regrettably forgotten afterwards.

Given \eqref{eq3.5} and \eqref{eq3.7}, a simple calculation provides the forward and backward drifts of the underlying diffusion process $X(\cdot)$. Preserving the notations $D_t X$ and $D^\ast_t X$ introduced after \eqref{eq2.23}, 
\begin{align}
&D_t X=\lim_{\Delta t\downarrow 0} E_t\left[\frac{X(t+\Delta t)-X(t)}{\Delta t}\right]=\hbar\nabla\log\eta_t(X)-A(X)\label{eq3.15}\\
&D^\ast_t X=\lim_{\Delta t\downarrow 0} E_t\left[\frac{X(t)-X(t-\Delta t)}{\Delta t}\right]=-\hbar\nabla\log\eta^\ast_t(X)-A(X)\qquad\tag{3.15*}\label{eq3.15*}\nonumber
\end{align}
In particular, since $P_t(dx)=\eta^\ast_t \eta_t(x)\,dx=\rho(x,t)\,dx$
\begin{equation}\label{eq3.16}
D^\ast_t X=D_tX-\hbar\nabla\log\rho.
\end{equation}
After substitution of \eqref{eq3.16} in the left hand side of \eqref{eq3.1} and an integration by part we obtain an elementary proof of this uncertainty relation \eqref{eq3.1}, justifying in this way the presence of two filtrations.

There is a new qualitative aspect in our probabilistic counterpart of Feynman's approach. Not surprisingly it comes from its boundary conditions.

The above construction of the processes, for a bounded below ``Hamiltonian" strongly suggests that natural boundary conditions should be two probability (densities) at the initial and final times:
\begin{equation}\label{eq3.17}
P_s(dx)=P_s(x)\,dx \mbox{ and } P_u(dz)=P_u(z)\,dz
\end{equation}
instead of the boundary conditions of the two adjoint equations \eqref{eq3.10} and \eqref{eq3.11}. But let us write the marginals of the joint probability $M_m$ \eqref{eq3.3}:
\begin{equation}\label{eq3.18}
\begin{cases}
\eta^\ast_s(x)\int h(s,x,u,z)\eta_u(z)\,dz=P_s(x)\\
\eta_u(z)\int \eta^\ast_s(x) h(s,x,u,z)\,dx=P_u(z)
\end{cases}
\end{equation}
in terms of the single integral kernel $h$. If $P_s$ and $P_u$ are arbitrarily given, Eq \eqref{eq3.18} is a non linear integral system for $(\eta^\ast_s,\eta_u)$ the two boundary conditions of the underlying adjoint PDE \eqref{eq3.10} and \eqref{eq3.11}, $s<t<u$.

Beurling has proved in 1960 the following general result: 

\begin{thm}[\cite{Beur60}]
Let the above integral kernel $h(s,x,u,z)=\left(e^{-\frac{1}{\hbar}(u-s)H}\right)$ $(x,z)$ be continuous, positive and defined on any locally compact space. Then the system \eqref{eq3.18} has a unique pair $(\eta^\ast_s,\eta_u)$ of positive, not necessarily integrable solutions, for any strictly positive probability densities $P_s(x), P_u(z)$.
\end{thm}

The proof of Beurling uses an entropic argument. This approach is quite natural in many respects (cf. \cite{Leonard} for instance) when handling this class of processes. Here, however, our present motivation comes from Mechanics and we shall not elaborate this Statistical Mechanics connection.

The above construction provides the solution of a stochastic boundary value problem, quite distinct from the Cauchy kind of problems originally inspired by Kolmogorov. Remarkably enough the processes solving such boundary value problems are necessarily time reversible (in a sense to be specified soon) although, as we will see, some of their partial characterizations reintroduce an ``arrow of time".

\vspace{0.5cm}

\noindent{\it Examples.}

\begin{enumerate}
\item Consider a Brownian $W_t$, $t\in\mathbb{R}^+$, on the real line, with diffusion coefficient $\hbar$ and initial probability density $\chi>0$. This is the case $A=V=0$ in \eqref{eq3.13}, i.e $H=-\frac{\hbar^2}{2}\Delta$. The traditional interpretation is that, given $\mu(dx)=\chi(x)\,dx$, $P^\mu(W_t\in dx)=\eta^\ast(x,t)$, where $\eta^\ast$ solves
\begin{equation}\label{eq3.19}
\begin{cases}
-\hbar\frac{\partial\eta^\ast}{\partial t}=H\eta^\ast\\
\eta^\ast(x,0)=\chi(x)
\end{cases}
\end{equation}

Now, if we wish to look at $W_t$ as a Bernstein reciprocal process $X(t)$, we should start from the same Hamiltonian $H$, a bounded time interval, say $I=[0,T]$ and the following boundary probability densities in \eqref{eq3.18}:

$P_0(x)=\chi(x)$ and $P_T(z)=\eta^\ast_\chi (x,T)$, where $\eta^\ast_\chi$ is the (positive) solution of Eq \eqref{eq3.19}. Of course, the kernel $h$ of \eqref{eq3.18} is the Gaussian one:
\begin{equation}\label{eq3.20}
h(0,x,T,z)=(2\pi\hbar T)^{-1/2}\exp -\frac{1}{2\hbar}\frac{|z-x|^2}{T}
\end{equation}
and the solution of Eq \eqref{eq3.18} (the ``Schr\"odinger system") on $[0,T]$ is trivial:
\begin{equation}\label{eq3.21}
\{\eta^\ast_0(x)=\chi(x),\eta_T(z)=1\}
\end{equation}
With those boundary conditions, the solutions of the two (heat) Eqs. \eqref{eq3.10} and \eqref{eq3.11} are clearly, $\forall \,t\in[0,T]$
\begin{align*}
\eta^\ast(x,t)=\eta^{\ast}_\chi(x,t),\quad \eta(x,t)=1.
\end{align*}
According to \eqref{eq3.15} the forward and backward drifts reduce therefore to 
\begin{align*}
D_t X=0,\; D^\ast_t X=-\hbar\nabla\log\eta^\ast_\chi(X,t)
\end{align*}
so that, denoting by $dX$ and $d_\ast X$, respectively, the It\^o differentials under $E_t$ in \eqref{eq3.15}, $X(t)$ solves both SDE
\begin{align}\label{eq3.22}
& dX(t)=\hbar^{1/2}dW_t,\nonumber\\ 
& d_\ast X(t)=-\hbar\nabla\log\eta^\ast_\chi(X(t),t)\,dt+\hbar^{1/2}d_\ast W^\ast_t
\end{align}
where $W^\ast_t$ denotes a Wiener process with respect to the filtration $\mathcal{F}_t$, $t\in [0,T]$.

With such choice of boundary probability densities $(P_0,P_T)$ it is hard, of course, to see any dynamical time symmetry. But let us switch them both, for the same kernel $h$ as before. Then $\{\hat{\eta}^\ast_0(x)=1, \hat{\eta}_T(x)=\chi(x)\}$ are also solutions of the system \eqref{eq3.18}. Indeed, $\hat{\eta}^\ast(x,t)=1$ and $\hat{\eta}(x,t)=\eta^\ast_\chi(x,T-t)$ solve the pair of heat equations \eqref{eq3.10}, \eqref{eq3.11}. The new associated process $\hat{X}(t)$, $t\in[0,T]$ is such that $D_t\hat{X}=\hbar\nabla\log\hat{\eta}(\hat{X},t)$ and $D^\ast_t\hat{X}=0$ and it is as well defined as the above diffusion $X(\cdot)$.

Notice that it follows easily from the above argument (or directly from the definitions \eqref{eq3.15} and \eqref{eq3.15*} that 
\begin{equation}\label{eq3.23}
D_t\hat{X}(t)=-D^\ast_t X(u+s-t),\quad s\leqslant t\leqslant u.
\end{equation}
This rule deforms the classical time reversal of derivatives into a more subtle one, involving necessarily two filtrations.

The full time symmetry of a Bernstein process (or equivalently of its probability measure) appears more clearly when considering processes not of independent increments.

\item For the same $H$ as in 1. pick the informal limiting case $P_s=\delta_x$ and $P_u=\delta_z$, corresponding to the solution $\eta^\ast_s(\cdot)=\delta_x$, $\eta_u(\cdot)=\delta_z$ of Eq \eqref{eq3.18}. So that $\eta^\ast(q,t)=h(s,x,q,t)$ and $\eta(q,t)=h(q,t,u,z)$ with $h$ as before. Eqs \eqref{eq3.15} and \eqref{eq3.15*} provide the two drifts 
\begin{equation}\label{eq3.24}
D_t X=\frac{z-X(t)}{u-t}, \quad D^\ast_t X=\frac{X(t)-x}{t-s}
\end{equation}
$X(t)$ is called the Brownian Bridge between $(s,x)$ and $(u,z)$. Defining $\hat{X}(t)=X(u+s-t)$, $s\leqslant t\leqslant u$, $\hat{X}(t)$ is another bridge traveling backward from $P_u=\delta_z$ to $P_s=\delta_x$.

\item The above construction is, in fact, independent of the form of the ``Hamiltonian" $H$. For instance, if one considers the (non-symmetric)
\begin{multline}\label{eq3.25}
H\eta(k)=U(k)\eta-c\nabla\eta-\frac{1}{2}\Delta\eta-\\\int_{\mathbb{R}^3}(\eta(k+y)-\eta(k)-y\nabla\eta(k) 1_{\{|y|\leqslant 1\}})\nu(dy)
\end{multline}
where $k$ represents a wave vector, in contrast with the space variable  $q$ of the Hamiltonians mentioned until now. This kind of non-local  Hamiltonian is relevant in the context of the "Momentum representation". In \eqref{eq3.25} $U:\mathbb{R}^3\to\mathbb{R}$ is continuous, bounded below, $c$ and $k\in\mathbb{R}^3$, $\nu(dy)$ a L\'evy measure on $\mathbb{R}^3\backslash\{0\}$. The resulting processes are well defined. They form an interesting  class of time reversible L\'evy processes (\cite{Priv04}, \cite{PrivZamb05}).
\end{enumerate}

The above examples suggest the following general notion of the time reversal involved here.

Any Bernstein process $X(t)$, $t\in[s,u]$ constructed as before, given a non necessarily symmetric Hamiltonian $H$ with integral kernel $h$ as in Beurling's Theorem, and any given pair of strictly positive probability densities $P_s(x)$ and $P_u(z)$ can be time reversed in the sense that $\hat{X}(t)=X(u+s-t)$, $t\in[s,u]$ is also a well defined process of the same class, evolving backward, and under $H^+$ from $P_u$ to $P_s$. This does not mean, of course, that the equations of motion of $X(t)$, to be discussed soon, exhibit invariance under time reversal. Already at the classical limit where $t\mapsto X(t)$ is a smooth trajectory and $P_s$, $P_u$ reduce to Dirac distributions the classical Lorentz Law (associated with $a\ne 0$ in Eq \eqref{eq2.1}) is not invariant under time reversal: the Lorentz force changes its sign. We shall come back on this.

We shall conclude this section by a comparison between Bernstein measures and the usual (``Euclidean") approach in Quantum Mechanics. Consider the definition \eqref{eq3.3} of the Markovian joint probability measure $M_m$. For $H$ self-adjoint and $\eta^\ast_s,\eta_u$ real-valued and bounded it can also be regarded as the $L^2$ scalar product 
\begin{equation*}
\langle\eta^\ast_s | e^{-\frac{1}{\hbar}(u-s)H}\eta_u\rangle
\end{equation*}
and expressed in terms of Wiener measure $\mu_w$ with density:
\begin{equation}\label{eq3.26}
\eta^\ast_s(\omega(s)) e^{-\frac{1}{\hbar}\int^u_s V(\omega(\tau))d\tau}\eta_u(\omega(u)) d\mu_w(\omega).
\end{equation}
When integrated over $\{\omega(s)\in A_s, \omega(u)\in A_u\}$ the expression coincides with \eqref{eq2.3}. This version of Feynman-Kac formula has been known and used since the sixties \cite{Nelson64}. A key difference with our construction is that, to produce Bernstein measures, $\eta^\ast_s$ and $\eta_u$ have first to be found as (positive) solutions of the system \eqref{eq3.18}, given any initial and final probability densities $P_s(dx)$ and $P_u(dz)$. Only then the reversibility of Bernstein measures and therefore their dynamical meaning will show up. Indeed, in Eq \eqref{eq3.26} the Wiener measure does not carry any specific dynamical meaning, in contrast with Bernstein measure, involving two drifts. This is already seen clearly in the Feynman-like formula (2.1) and, of course, in the equations of motion that we are going to obtain in the next section.

\section{Stochastic Dynamics and Symmetries}
\setcounter{equation}{0}

There are two approaches to classical dynamics (or the classical calculus of variations) the Lagrangian and the Hamiltonian one. Feynman's method suggests to start from the former one. The classical system with Lagrangian \eqref{eq2.11} will be our guide. We already know that, in relation with the $\mathcal{P}_t$ filtration, the classical $\dot{\omega}$ should become $D_t X$ (apart from its imaginary unit factor) under our stochastic deformation. So, from the transformations used in \eqref{eq3.12}, \eqref{eq3.13}  the Lagrangian should be proportional to 
\begin{equation}\label{eq4.1}
\mathcal{L}(X, D_t X)=\frac{1}{2}|D_t X|^2+V(X)+A\cdot D_t X+\frac{\hbar}{2}\nabla\cdot A.
\end{equation}
The last term of \eqref{eq4.1} requires some explanation , useful also for later purposes. 

Three definitions of stochastic integrals will be necessary in our time-symmetric context, when looking for the stochastic deformation of classical expressions of the form
\begin{equation}\label{eq4.2}
\int^u_t y(\omega) d\omega(\tau)
\end{equation}
in the Action functional. In It\^o's theory, $y$ can be a very general functional. Here we shall need only $y(\omega)(\tau)$ of the form $A(\omega(\tau))$ with $A:\mathbb{R}^3\to\mathbb{R}^3$ and such that $E\int^u_t |A|^2(X(\tau))d\tau<\infty$. Then, with respect to the increasing filtration $\mathcal{P}_\tau$,
\begin{equation*}
\int^u_t A(X)\, dX(\tau)=\lip_{\substack{\max|\tau_j-\tau_{j-1}|\to 0\\1\leqslant j\leqslant N}}\sum^N_{j=1} A(X(\tau_{j-1}))(X(\tau_j)-X(\tau_{j-1}))
\end{equation*}
where l.i.p means limit in probability. Using \eqref{eq3.15},
\begin{equation}\label{eq4.2*}
E\int^u_t A\, dX(\tau)=E\int^u_t A\,D_\tau X d\tau.
\end{equation}
With respect to the decreasing filtration $\mathcal{F}_\tau$ and introducing the notation $d_\ast X(\tau)$ for the backward differential involved in \eqref{eq3.15*},
\begin{equation*}
\int^u_t A(X) d_\ast X(\tau)=\lip_{\substack{\max|\tau_j-\tau_{j-1}|\to 0\\1\leqslant j\leqslant N}}\sum^N_{j=1} A(X(\tau_j))(X(\tau_j)-X(\tau_{j-1}))
\end{equation*}
and therefore, 
\begin{equation}\label{eq4.3}
E\int^u_t A d_\ast X(\tau)=E\int^u_t A D^\ast_\tau X d\tau.
\end{equation}
The third one is due to Stratonovich: (when, in addition, $E\int^u_t |\nabla A|^2(X(\tau))d\tau<\infty$)
\begin{multline*}
\int^u_t A\circ dX(\tau)=\lip_{\substack{\max|\tau_j-\tau_{j-1}|\to 0\\1\leqslant j\leqslant N}}\sum^N_{j=1}\frac{1}{2}[A(X(\tau_{j-1}))+\\A(X(\tau_j))](X(\tau_j)-X(\tau_{j-1}))
\end{multline*}
and 
\begin{equation}\label{eq4.4}
E\int^u_t A\circ dX(\tau)=E\int^u_t A\cdot \frac{1}{2}(D_\tau X + D^\ast_\tau X)d\tau.
\end{equation}
It follows from the usual (forward) It\^o calculus \cite{Ikeda81} that 
\begin{equation}\label{eq4.5}
A\circ dX(\tau)=A\cdot dX(\tau)+\frac{\hbar}{2}\nabla\cdot A\, d\tau
\end{equation}
Together with \eqref{eq4.2} this justifies the following definition of the Action functional relevant for the dynamics of our system with Lagrangian \eqref{eq4.1}:
\begin{multline}\label{eq4.6}
J[X]=E_{xt}\int^u_t\mathcal{L}(X(\tau), D_\tau X(\tau))\,d\tau\\
=E_{xt}\left\{\int^u_t \frac{1}{2} |D_\tau X(\tau)|^2+V(X(\tau))\,d\tau +\int^u_t A\circ dX(\tau)\right\}
\end{multline}
where $E_{xt}$ denotes the conditional expectation given $X(t)=x$, $t<u$.

What about the geometrical meaning of $J$ ? Let us define, for any (hyperregular \cite{Mars94}) Lagrangian $\mathcal{L}$, the (forward) Momentum as classically by 
\begin{equation}\label{eq4.7}
P=\frac{\partial \mathcal{L}}{\partial D_t X} (X,D_t X)
\end{equation} 
assuming that Eq \eqref{eq4.7} is solvable in $D_t X=\phi(P,X)$. This is the case for the system \eqref{eq4.1} and 
\begin{equation}\label{eq4.8}
D_t X=P-A.
\end{equation}
Let us recall that the Poincar\'e-Cartan  1-form $\omega_{PC}$ of a classical system with Hamiltonian $h(q,p)$ is defined by $pdq-hd\tau$ on the extended phase space. Its section on the $(q,\tau)$ submanifold is often denoted  by  $\tilde{\omega}=p(q,\tau)\,dq-h(q,p(q,\tau))\,d\tau$.

The deformation  of these notions allows us to write our Action \eqref{eq3.7} as 
\begin{equation}\label{eq4.9}
J[X]=E_{xt}\int^u_t\tilde{\omega}_{PC}=E_{xt}\int^u_t P\circ dX(\tau)-h(X(\tau),\tau)\,d\tau
\end{equation}
where $h$ is now a scalar field, called the energy function. Notice that the drift is regarded as a function of $X$: $D_t X=B(X,t)$, as in \eqref{eq3.15}. For \eqref{eq4.1} we find 
\begin{equation}\label{eq4.10}
h(X,\tau)=\frac{1}{2}|B|^2+\frac{\hbar}{2}\nabla\cdot B-V(X)
\end{equation}
We shall come back later on the meaning of $h$.

Assume now, following Feynman, that we are only given the Lagrangian $\mathcal{L}$ of \eqref{eq4.6}. For the sake of generality we shall add a smooth (final) boundary condition: $J[X]=E_{xt}\left\{\int^u_t\mathcal{L}\, d\tau+S_u(X(u))\right\}$. Such an addition is well known in classical calculus of variations (and optimal control theory) as a Bolza variational problem.

When not referring to the $\mathcal{L}$ of \eqref{eq4.1} we shall assume from now on that $\mathcal{L}, S_u$ are continuous and that, for some constants $c,k,|\mathcal{L}|\leqslant c(1+|X|^k)+|DX|^k, |S_u(X)|\leqslant c(1+|X|^k)$.\\

How can we characterize critical points of $J$ ?

We shall pick, as domain $\mathcal{D}_J$ of $J$ the set of diffusions $X$ absolutely continuous with respect to the Wiener measure $P^{\hbar}_W$ with diffusion matrix $\hbar I$ ($I$ the $3\times 3$ Identity matrix), and arbitrary Borel measurable drift $B$ of $\mathbb{R}^3\times[s,u]$ into $\mathbb{R}^3$.

\begin{defn}
Such a process $X$ is extremal for $J$ if 
\begin{equation}\label{eq4.11}
E_{xt}[\nabla J[X](\delta X)]=E_{xt}\left[\lim_{\varepsilon\to 0}\frac{J[X+\varepsilon\delta X]-J[X]}{\varepsilon}\right]=0,
\end{equation}
for any variation $\delta X$ in  the Cameron-Martin space $C_{xt}$ preserving the absolute continuity under the shift $X+\varepsilon\delta X$ (cf. \cite{Cruz91}, \cite{Chung03}, \cite{Mall97}).
\end{defn}

Then 
\begin{multline*}
0=E_{xt}[\nabla J[X](\delta X)]=E_{xt}\int^u_t \left(\frac{\partial\mathcal{L}}{\partial X} \delta X+\frac{\partial\mathcal{L}}{\partial D_\tau X} D_\tau \delta X\right)\,d\tau+\\
E_{xt}[\nabla S_u(X(u))\delta X(u)].
\end{multline*}
In the second term, we preserve the notation $D_\tau$, this time for the (extended) infinitesimal generator of $X\in \mathcal{D}_J$. This one is defined, for any smooth enough function $f$ by
\begin{multline}\label{eq4.12}
D_\tau f(X(\tau),\tau)=\lim_{\Delta\tau\downarrow 0} E_\tau\left[\frac{f(X(\tau+\Delta\tau),\tau+\Delta\tau)-f(X(\tau),\tau)}{\Delta\tau}\right]\\
=\left(\frac{\partial}{\partial\tau}+B\cdot\nabla+\frac{\hbar}{2}\Delta\right) \, f(X(\tau),\tau)
\end{multline}
When applied, in particular, to $X$ itself, $D_\tau X=B(X,\tau)$, as in \eqref{eq3.15}. Under $E_{xt}[\dots]$, $D_\tau$ satisfies, by It\^o formula, an integration by parts formula. Since, here,  $\delta X(\tau)$ is of bounded variation, there is no (It\^o's) extra term in this formula:
\begin{multline*}
0=E_{xt}\int^u_t \left(\frac{\partial\mathcal{L}}{\partial X}-D_\tau\left(\frac{\partial\mathcal{L}}{\partial D_\tau X}\right)\right) \delta X(\tau)\, d\tau+\\
E_{xt}\left[\left(\frac{\partial \mathcal{L}}{\partial D_\tau X}+\nabla S_u\right)(X(u))\delta X(u)\right]
\end{multline*}
To conclude we need a stochastic deformation of (Lagrange) fundamental lemma of the classical calculus of variations.

For any $\phi(X)(\tau)\in C_{xt}$, the Cameron-Martin space corresponding to the space $\Omega_{xt}$ of continuous paths starting from $x$ at time $t$, if 
\begin{equation*}
E_{xt}\int^u_t \phi(X)(\tau)\delta X(\tau)\, d\tau=0,\; \forall\,\delta X\text{ in } C_{xt} \text{ then,}
\end{equation*}
by orthogonality in the Hilbert space $C_{xt}$, the conclusion $\phi(X)(\tau)=0$ a.s, $\forall\,\tau$, is true. When $\phi(X)$ is not of bounded variations but, as in our case, of the form $F(X(\tau),\tau)$, for $F$ real valued, continuous in both variables and in $L^2(\Omega_{xt} \times[t,u])$ i.e 
\begin{equation*}
E_{xt}\int^u_t |F(X(\tau),\tau)|^2 d\tau <\infty,
\end{equation*}
then, using a sequence of $\phi_n(z(\tau),\tau)\in C_{xt}$ converging to $F$ in $L^2(\Omega \times[t,u])$,we conclude that, a.s, $F(X(\tau),\tau)=0$ for almost every $\tau$. By continuity $F(X(\tau),\tau)=0,\quad \forall\tau$.

\subsection*{Stochastic Euler-Lagrange Theorem}
A necessary condition for a diffusion $X$ to be extremal for $J[X]$ as before is that the following stochastic (almost sure) Euler-Lagrange equation is satisfied, as well as the final transversality  condition
\begin{equation}\label{eq4.13}
\textrm{(SEL)\quad}
\begin{cases}
D_\tau\left(\frac{\partial\mathcal{L}}{\partial D_\tau X}\right)-\frac{\partial\mathcal{L}}{\partial X}=0, & t<\tau<u\\
\\
\frac{\partial\mathcal{L}}{\partial D_\tau X}(X(u),D_\tau X(u))=-\nabla S_u(X(u)), & X(t)=x.
\end{cases}
\end{equation}
For instance, when $\mathcal{L}$ is as in \eqref{eq4.1}, (SEL) reduces to 
\begin{equation}\label{eq4.14}
\begin{cases}
D_\tau D_\tau X(\tau)=\nabla V(X(\tau))+D_\tau X(\tau)\wedge\rot A+\frac{\hbar}{2}\rot\rot A\\
D_\tau X(u)+A(X(u))=-\nabla S_u(X(u)).
\end{cases}
\end{equation}
One should notice the $\hbar$ dependent term, in addition to the Lorentz force, on the r.h.s of Eq \eqref{eq4.14}. It is natural to interpret (SEL) as the $(\mathcal{P}_t)$ stochastic deformation of its classical counterpart.

Now we can come back on Burdzy's ``dynamical paradox" (end of \S2). The Brownian motion  used there is, for us, critical point of the Action \eqref{eq4.6} with $V=A=0$ and final boundary condition $S_u=0$ (cf. remark after \eqref{eq4.10}). In this case SEL \eqref{eq4.13} reduces to 
\begin{equation*}
(\ast)
\begin{cases}
D_\tau D_\tau X(\tau)=0, \quad 0<\tau<u\\
X(0)=0,\quad D_\tau X(u)=0
\end{cases}
\end{equation*}

Our Action involves  the same classical free Lagrangian as in Burdzy's remark; but evaluated now among well defined (forward) drifts. The left hand side of the almost sure equation $(\ast)$ provides the probabilistic counterpart of the ODE characterizing the most likely shape of Brownian path. But Eq $(\ast)$ shows more, regardless of associated boundary conditions. The drift $D_\tau X$ of such a system, stochastic deformation of the momentum of a classical  free particle, must be a $\mathcal{P}_\tau$ martingale. Let us recall, indeed, that a stochastic process $M_t$ defined on our basic probability space $(\Omega,\sigma, \{\mathcal{P}_t\},\{\mathcal{F}_t\})$ is called a $\mathcal{P}_t$-martingale if it is $\mathcal{P}_t$ measurable $\forall\,t$, $E[|M_t|]<\infty$ $\forall\,t$ and $E[M_t|\mathcal{P}_s]=M_s$ $\forall\, t\geqslant s$. In a Markovian case like $(\ast)$ and when $\mathcal{P}_s$ is the filtration generated by a process, here $X(\tau)=W(\tau)$, $0<\tau\leqslant s$, $E[M_t|\mathcal{P}_s]$ reduces to $E_s[M_t]$. Then it follows from the definition \eqref{eq4.12} of $D_\tau$ applied to $M_\tau$ that $D_\tau M_\tau=0$. Reciprocally, any $M_\tau$ in the domain of $D_\tau$ satisfying this equation is a $\mathcal{P}_\tau$-martingale. For the special boundary conditions of Eq $(\ast)$, $X(\tau)=W(\tau)$ and $D_\tau X(\tau)$ is a trivial martingale, the constant $0$. Since, in our time symmetric framework another, decreasing, filtration $\mathcal{F}_\tau$ is also available, one can as well define backward martingales $M^\ast_\tau$ by the condition $D^\ast_\tau M^\ast_\tau=0$.

The central role of the concept of martingale (introduced by P. L\'evy, 1934) in Stochastic Analysis, has been particularly elaborated by J.L. Doob. The traditional interpretation of a martingale $M_t$ is that it modelizes a fair game. The knowledge of the past information $\mathcal{P}_s$ until time $s$ does not help us to assess the expectation of $M_t$: it will be $M_s$. This justifies the non existence of winning strategy in a fair game, even if we have a recording of its whole history.

Resisting  a financial interpretation, it is clear that a martingale is, in our context, the closest possible analogy with a conserved quantity (a ``first integral") of a classical dynamical system. The strict conservation in time is lost, but the absolute expectation of the martingale $M_t$ remains constant. This follows from  taking the expectation of its definition $E[M_t|\mathcal{P}_s]=M_s$.

Regarding the Hamiltonian approach of the dynamics, one can define, as classically, using the hypothesis after \eqref{eq4.7},
\begin{equation}\label{eq4.15}
\mathcal{H}(X,P)=P\phi(P,X)-\mathcal{L}(X,\phi(P,X)).
\end{equation}
The deformed version of the Hamiltonian differential equations becomes the almost sure 

\subsection*{Stochastic Hamiltonian equations: }

\begin{equation}\label{eq4.16}
\begin{cases}
D_\tau X=\frac{\partial\mathcal{H}}{\partial P}\\
D_\tau P=-\frac{\partial\mathcal{H}}{\partial X}.
\end{cases}
\end{equation}
For instance when $\mathcal{L}$ is the one of \eqref{eq4.1}, $\mathcal{H}$ reduces to
\begin{equation}\label{eq4.17}
\mathcal{H}(X,P)=\frac{1}{2}|P-A(X)|^2-V(X)-\frac{\hbar}{2}\nabla\cdot A
\end{equation}
and the Hamiltonian equations become
\begin{equation}\label{eq4.18} \textrm{(SHE)\quad}
\begin{cases}
D_\tau X=P-A\\
D_\tau P=(D_\tau X\cdot\nabla) A+D_\tau X\wedge\rot A+\frac{\hbar}{2}\rot\rot A+\frac{\hbar}{2}\Delta A+\nabla V.
\end{cases}
\end{equation}
The first equation reduces to another version of \eqref{eq3.15}. Its substitution in the second one is consistent with the Euler-Lagrange equation \eqref{eq4.14}.

Almost sure equations like SEL \eqref{eq4.13} and SHE \eqref{eq4.16} may seem to be odd but, in fact, they were (rather deeply) hidden behind some classical results of Stochastic Control Theory found around the eighties \cite{Bismuth,Flem06, Hauss81} from a very different viewpoint. In this context, it is the classical theory of Characteristics which was stochastically deformed.We shall give here only a hint about this connection, particularly natural in our Path Integral perspective, for the simple case above where $A=0$.

Consider, then, the Action functional $J[X]$, where the drift $B$ of the critical process (with fixed diffusion matrix) we are looking for, called the ``control" is just supposed to be $\mathcal{P}_t$-measurable and such that $E\int^u_t|B(\tau)|^n d\tau<\infty$, $n\in\mathbb{N}$. Notice that this include, now, non Markovian processes, a natural hypothesis in our perspective.

Consider any scalar field $S$ in the domain of the infinitesimal generator $A^{B(\tau)}=\frac{\partial}{\partial\tau}+B(\tau)\nabla+\frac{\hbar}{2}\Delta$ of such a process $X(\tau)$, with
\begin{align*}
&E_{xt}|S(X(u),u)|<\infty
\intertext{and }
&E_{xt}\int^{u}_{t}|A^{B(\tau)}S(X_\tau,\tau)|\,d\tau<\infty
\intertext{and such that Dynkin formula holds:}
&E_{xt}\int^{u}_{t} A^{B(\tau)}S(X_\tau, \tau)\, d\tau=E_{xt} S(X(u),u)-S(x,t).
\end{align*}
So, for any $X(\tau)$ in the above class, and $\mathcal{L}(X,D_\tau X)=\frac{1}{2}|D_\tau X|^2+V(X)$
\begin{equation*}
J[X]=E_{xt}\int^u_t \mathcal{L}(X(\tau), B(\tau))\,d\tau+E_{xt}S_u(X(u))
\end{equation*}

\begin{thm} [\cite{Flem06}]
Let $S(x,t)$be a classical solution of the deformed Hamilton-Jacobi equation (known as ``Hamilton-Jacobi Bellmann", or HJB)
\begin{equation}\label{eq4.19}
\begin{cases}
\frac{\partial S}{\partial t}-\frac{1}{2}|\nabla S|^2+\frac{\hbar}{2}\Delta S+V=0\\
S(x,u)=S_u(x).
\end{cases}
\end{equation}
Then $S(x,t)\leqslant J[X]$, $\forall X$ in the above class. Moreover, for 
\begin{equation}\label{eq4.20}
B(t)=B(x,t)=-\nabla S(x,t)
\end{equation}
this inequality becomes an equality.
\end{thm}

The above class of drifts $B(\tau)$ is called the one of admissible (progressively measurable) control processes by Fleming and Soner \cite{Flem06}. As said before, Markov property is not, a priori, natural in our context. But if we already know that the extremal process will be Markovian and, therefore, its drift a function $B(X(t),t)$ and we are looking, like here, for a classical solution of the ``dynamic programming equation" \eqref{eq4.19} then \eqref{eq4.20} is an ``optimal Markov control policy". Dynkin formula becomes a consequence of It\^o formula, for $S$ continuous, regular and whose partial derivatives satisfy polynomial growth conditions.

To understand the relation with our construction (for $A=0$), define 
\begin{equation}\label{eq4.21}
\eta(x,t)=e^{-\frac{1}{\hbar}S(x,t)}
\end{equation}
then HJB reduces to Eq \eqref{eq3.11} with a positive final condition,
\begin{equation}\label{eq4.22}
\hbar\frac{\partial \eta}{\partial t}=-\frac{\hbar^2}{2}\Delta \eta+V, \quad \eta(x,u)=e^{-\frac{1}{\hbar}S_u(x)}
\end{equation}
and the extremal, indeed minimal, diffusion of $J[X]$ is our Markovian (forward) Bernstein process of drift \eqref{eq3.15} $D_t X=B(X,t)=\hbar\nabla\log\eta(x,t)$.

The relation \eqref{eq4.21} is sometimes called ``Fleming logarithmic transformation" (\cite{FleRis75}). It is a remarkable coincidence (?) that the origin of this transformation goes back to Schr\"odinger's paper of 1926 where he introduced his equation \eqref{eq2.14}. Indeed, his publication \cite{Schr32}, written 5 years later, is also the origin of our program of Stochastic deformation.

It is in fact possible to prove in a purely geometric way that the extremal points of $J[X]$ are minimal \cite{Zamb}.

The hypothesis, in the last Theorem, that $S$ is a classical solution of HJB is, of course, much too restrictive. The construction holds under weaker conditions (cf. \cite{Vuil12}). Regarding HJB, the appropriate notion of weak solution is the one of viscosity solution (cf. \cite{Flem06}).

When $\hbar=0$, the existence of a classical, global solution of Hamilton-Jacobi equation is a condition of Complete Integrability of the system and, if available, the gradient of this equation coincides with the second Hamiltonian equation, $\frac{dp}{dt}=-\frac{\partial \mathcal{H}}{\partial x}=-\nabla V(x)$, when $\mathcal{H}(x,p)=\frac{1}{2} p^2 +V(x)$.

The stochastic deformation of this integrability condition is that, the gradient of \eqref{eq4.19}, using $D_t X=\hbar\nabla\log\eta_t$ (i.e \eqref{eq3.15} for $A=0$) reduces to, almost surely, $D_t D_t X(t)=\nabla V(X(t))$ namely \eqref{eq4.14} or, equivalently, the second deformed Hamiltonian equation \eqref{eq4.16} in the same special case (cf. \cite{Zamb09}).

Coming back to the deformed Hamiltonian $\mathcal{H}$ of \eqref{eq4.17} one observes that it does not coincide with the energy function $h$ \eqref{eq4.10} of the Poincar\'e-Cartan 1-form. So, what is the meaning, if any, of $h(X(\tau),\tau)$ ?

A key observation is that $D_\tau h(X(\tau),\tau)=0$, for the critical, i.e dynamical, diffusion $X(t)$, with generator 
\begin{equation}\label{eq4.23}
D_\tau=\frac{\partial}{\partial\tau}+\hbar\nabla\log\eta\nabla+\frac{\hbar}{2}\Delta.
\end{equation}
In other words, $h(X(\tau),\tau)$ is a $\mathcal{P}_\tau$-martingale, a natural deformation of the classical notion of constant of the motion or first integral. Is it accidental ?

The answer is negative. Before mentioning the Theorem showing that all first integrals of our stochastic dynamical system are indeed martingales, let us stress that Feynman could not find such a result in his informal time discretized account of quantum dynamics.

As it is clear from \eqref{eq3.15}, the processes extremal for $J$ are entirely built from (positive) solutions of a parabolic equation \eqref{eq3.11} and the coefficients of its Hamiltonian. We shall stick to the simple case $A=0$, i.e Eq \eqref{eq4.22}.

Consider this equation, written now as 
\begin{equation}\label{eq4.24}
\hat{H}\eta=\left(\hbar\frac{\partial}{\partial t}+\frac{\hbar^2}{2}\Delta-V\right)\eta=0
\end{equation}
and define a linear differential operator of the form
\begin{equation}\label{eq4.24'}
N=T(t)\frac{\partial}{\partial t}+Q_i(x,t)\frac{\partial}{\partial x_i}-\frac{1}{\hbar}\phi(x,t)
\end{equation}
where the summation convention is used, for $1\leqslant i\leqslant 3$ and the (unknown) coefficient $T,Q,\phi$ are analytic in $x$ and $t$. Then $N$ is infinitesimal generator of a Lie symmetry group of Eq \eqref{eq4.24} if 
\begin{equation}\label{eq4.25}
\hat{H}\eta=0\Rightarrow \hat{H}N\eta=0.
\end{equation}
Lie proved long ago that this is the case if $(T,Q,\phi)$ solve the following ``determining equations": 
\begin{equation}\label{eq4.26}
\begin{cases}
\frac{dT}{dt}=2\frac{\partial Q_i}{\partial x_i} & \frac{\partial Q_i}{\partial x_j}+\frac{\partial Q_j}{\partial x_i}=0, \quad i=1,2,3,\quad j\ne i\\
\frac{\partial Q_i}{\partial t}=\frac{\partial\phi}{\partial x^i}\\
\frac{\partial\phi}{\partial t}+\frac{\hbar}{2}\Delta\phi=\frac{dT}{dt} V+Q_i\frac{\partial V}{\partial x_i}+T\frac{\partial V}{\partial t}
\end{cases}
\end{equation}

The (local) Lie symmetry group of Eq \eqref{eq4.22} results from product of exponentials of such operators $N$.

When $x\in\mathbb{R}^3$ and $V=A=0$, in \eqref{eq3.13} for instance, the symmetry group is 13 dimensional \cite{Olver86}.

Now consider a classical Lagrangian $L(\omega,\dot{\omega},t)$. One general version of the classical Theorem of Noether is the following:
\begin{equation*}
v=T(t)\frac{\partial}{\partial t}+Q_i(\omega,t)\frac{\partial}{\partial{\omega_i}}+\left(\frac{dQ_i}{dt}-\dot{\omega}_i\frac{dT}{dt}\right)\frac{\partial}{\partial \dot{\omega}_i}
\end{equation*}
is called a divergence symmetry of $L$ if there is a scalar field $\phi=\phi(\omega,t)$ such that
\begin{equation}\label{eq4.27}
v(L)+L\frac{dT}{dt}=\frac{d\phi}{dt}.
\end{equation}
When $L$ admits such a divergence symmetry, then along each extremal of the classical action $S_L$,
\begin{equation}
\frac{d}{dt}\left[\frac{\partial L}{\partial\dot{\omega}_i} Q_i-\left(\frac{\partial L}{\partial\dot{\omega}_i}\dot{\omega}_i -L\right) T-\phi\right]=\label{eq4.28}
\end{equation}
Equivalently, the expression between brackets is a constant of the motion of the system. In particular, for the elementary Lagrangian associated with the Hamiltonian of Eq \eqref{eq4.24}, this means that
\begin{equation*}
\frac{d}{dt}\left[\dot{\omega}_i Q_i - \left(\frac{1}{2}|\dot{\omega}|^2+V(\omega)\right)T-\phi\right]=0.
\end{equation*}

For instance, any conservative system (i.e with $V$ time independent) admits $v=\frac{\partial}{\partial t}$ i.e $T=1$, $Q=0$, $\phi=0$ and \eqref{eq4.28} reduces to the energy conservation.

The stochastic deformation of this Theorem, for the same class of elementary systems is the 

\subsection*{Stochastic Noether's Theorem {\rm\cite{Thieu97}}}

If the Lagrangian $\mathcal{L}(X,D_t X,t)=\frac{1}{2}|D_t X|^2+V(X,t)$ admits a divergence symmetry of the form
\begin{equation}\label{eqref4.29}
T(t)\frac{\partial}{\partial t}\mathcal{L}+Q_i\frac{\partial}{\partial X_i}\mathcal{L}+
\left(D_t Q_i-D_t X_i\frac{dT}{dt}\right)\frac{\partial\mathcal{L}}{\partial D_t X_i}+\mathcal{L}\frac{dT}{dt}=D_t\phi
\end{equation}
for any analytic $T,Q,\phi$ solving the ``Determining equations" \eqref{eq4.26} then along any Bernstein diffusion $X(\cdot)$ extremal for the action $J[X]$, almost surely 
\begin{equation}\label{eq4.30}
D_t(D_t X_i Q_i - hT-\phi)(X(t),t)=0
\end{equation}
where $h$ is the energy function \eqref{eq4.10}.

For instance, if $V$ is time independent, $\mathcal{L}$ admits $T=1$, $Q=\phi=0$ and we recover our energy martingale.

The role of the ``divergence" term $\phi$ is deep, both in its classical and deformed sense. $D_t\phi$ is the stochastic deformation of the classical notion of ``Null Lagrangian", whose associated Euler-Lagrange equation is trivially satisfied. Divergence symmetries  are not exceptional, even classically. For instance, for the Lagrangian of the classical n-body system, we have only divergence symmetry under Galilean transformations. Indeed, velocity translations are admissible but $\phi$ is nonzero. The same remains true after stochastic deformation.

The stochastic Noether Theorem is a theorem of structure, here, without which our deformation would be dynamically meaningless.

But let us observe that, from the start of this Section only one filtration, the increasing one $\mathcal{P}_t, s<t<u$, has been used. As a result of this, the stochastic Euler-Lagrange equation \eqref{eq4.14}, for instance, is certainly not invariant under time reversal in the sense defined in Section 3. So \eqref{eq4.14} cannot be the full dynamical characterization of processes respecting, by construction, this invariance.

But it is easy to find the solution of the puzzle. On the time interval $[s,u]$, the time reversal of the Action functional \eqref{eq4.6} is 
\begin{equation*}\label{eq4.7*}
E^{xt}\int^t_s\left(\frac{1}{2}|D^\ast_\tau X(\tau)|^2+V(X(\tau))\right)\,d\tau+\int^t_s A\circ dX(\tau)\tag{4.7*}
\end{equation*}
where we have adopted the notation $E^{xt}$ for a conditional expectation given the future configuration  $X(t)=x$, $s<t$, and used the rule \eqref{eq3.23}.

Calling $J^\ast [X]$ the functional \eqref{eq4.7*}, we can look for its critical points among diffusions with fixed diffusion matrix, solving a backward (i.e $\mathcal{F}_t$) stochastic differential equation whose (backward) drift is unknown. Notice that, now, the Stratonovich integral in \eqref{eq4.7*} must be interpreted using the backward version of the relation \eqref{eq4.5} namely, according to It\^o \cite{Ito78},
\begin{equation*}\label{eq4.5*}
A\circ d X(\tau)=A\,d_\ast X(\tau)-\frac{\hbar}{2}\nabla\cdot A\, d\tau.\tag{4.6*}
\end{equation*}
This means that, with respect to $\mathcal{F}_\tau$, the Lagrangian of $J^\ast$ is now represented by 
\begin{multline*}\label{eq4.1*}
\mathcal{L}^\ast(X(\tau), D^\ast_\tau X(\tau))=\frac{1}{2}|D^\ast_\tau X(\tau)|^2 + V(X(\tau))+\\
A(X(\tau)) D^\ast_\tau X(\tau)-\frac{\hbar}{2} \nabla\cdot A. \tag{4.1*}
\end{multline*}
Then, one checks in the same way as before, that the extremal point of $J^\ast[X]$, in fact a minimum, is unique and that its backward drift is given in term of a positive solution of Eq \eqref{eq3.10} by the expression:
 \begin{equation*}
 D^\ast_\tau X=-\hbar\nabla\log\eta^\ast_t(X)-A(X).
 \end{equation*}
 The (backward) stochastic Euler-Lagrange equation it solves (ignoring boundary condition at $t>s$), is
 \begin{equation*}\label{eq4.14*}
 D^\ast_\tau D^\ast_\tau X(\tau)=\nabla V(X(\tau))+D^\ast_\tau X(\tau)\wedge\rot A-\frac{\hbar}{2}\rot\rot A.\tag{4.15*}
 \end{equation*}
 As a matter of fact, such a calculation is not even necessary. Indeed, as said before (cf \eqref{eq3.23}), $D_\tau\to-D^\ast_\tau$ and $A\to -A$ under time reversal. This means that both \eqref{eq3.15*} and \eqref{eq4.14*} are time reversed versions of their forward counterparts \eqref{eq3.15} and \eqref{eq4.14}.
 
Since, in particular, \eqref{eq4.14} and \eqref{eq4.14*} provide different informations, associated with $\mathcal{P}_t$ and $\mathcal{F}_t$ respectively, about the same Bernstein diffusion, the complete, time-symmetric, dynamical equation of $X(\tau)$, $s\leqslant \tau\leqslant u$, is 
 \begin{multline}\label{eq4.31}
 \frac{1}{2}(D_\tau D_\tau X(\tau)+ D^\ast_\tau D^\ast_\tau X(\tau))=\\
 \frac{1}{2}(D_\tau X(\tau)+D^\ast_\tau X(\tau))\wedge\rot A(X(\tau))+\nabla V(X(\tau)).
 \end{multline}
Let us stress that, now, this stochastic deformation of the classical Euler-Lagrange equation in an electromagnetic field:
\begin{equation*}
\ddot{\omega}(\tau)=\dot{\omega}(\tau)\wedge\rot A(\omega(\tau))+\nabla V(\omega(\tau))
\end{equation*}
involving the deformed Lorentz force on the right hand side, is indeed invariant under time reversal, as it should. Using the relations , for $f\in C^2$, 
\begin{equation*}
\frac{d}{d\tau} E[f(X(\tau))]=E[D_\tau f(X(\tau))]=E[D^\ast_\tau f(X(\tau))]
\end{equation*}
following, for instance, from Dynkin formula (cf. also \cite{Nelson67}), we can get even closer to Feynman dynamical law \eqref{eq2.18} in taking the absolute expectation of \eqref{eq4.31}:
\begin{equation}\label{eq4.34}
\frac{d^2}{d\tau^2} E[X(\tau)]=E\left[\frac{1}{2}(D_\tau X(\tau)+D^\ast_\tau X(\tau))\wedge\rot A(X(\tau))+\nabla V(X(\tau))\right].
\end{equation}
Notice also that what plays the role of the time derivative $\dot{\omega}(\tau)$ in Feynman's law of motion \eqref{eq2.18} is now the average of the two drifts. This average changes correctly its sign under time reversal in contrast with each of the drifts taken separately. Regarding the specific symmetric form of second time derivative on the l.h.s. of \eqref{eq4.31}, which is the rigorous version of Feynman's discretization in Eq. \eqref{eq2.18}, it is worth observing that it was mentioned in \cite{Nelson67} as a possible definition of acceleration. Unfortunately, it was not identified as the proper one for an ``Euclidean" stochastic deformation of classical mechanics. In fact, the existence of such a theory was only discovered twenty years after (in \cite{Zamb86}).

Using the same method, it is easy to find the backward version of our Stochastic Noether Theorem, for instance, producing backward martingales of the system.

\section{Computational and Geometric content}
\setcounter{equation}{0}

Let us start with some consequences of our Noether Theorem.

Although it is clear, for a member of the community of Geometric Mechanics (in particular) that Noether Theorem is the key to start a serious study of dynamics, we shall try to show its interest also for the theory of stochastic processes itself. 

Consider diffusions on the line, for simplicity, with $A=V=0$ in Eq \eqref{eq3.13} or, equivalently a Lagrangian \eqref{eq4.1} reduced to
\begin{equation}\label{eq5.1}
\mathcal{L}(X,D_t X)=\frac{1}{2} |D_t X|^2.
\end{equation}
One verifies that $T=2t$, $Q=x$, $\phi=0$ solves the one dimensional version of the determining equations \eqref{eq4.26} for $V=0$. So $N=2t\frac{\partial}{\partial t}+x\frac{\partial}{\partial x}$ generates a one-parameter symmetry group:
\begin{equation}\label{eq5.2}
(e^{\alpha N})(t,x,\eta)=(e^\alpha x, e^{2\alpha} t,\eta)=(t_\alpha,x_\alpha,\eta_\alpha).
\end{equation}
This implies that if $\eta=\eta(x,t)>0$ solves the free heat equation \eqref{eq4.22} for $V=0$, i.e $\hbar\frac{\partial\eta}{\partial t}=-\frac{\hbar^2}{2}\frac{\partial^2\eta}{\partial x^2}$, so does
\begin{equation}\label{eq5.3}
\eta_\alpha(x,t)=e^{-\alpha N}\eta=\eta(e^{-\alpha x}x,e^{-2\alpha}t).
\end{equation}
Then define $h_\alpha(x,t)=\frac{\eta_\alpha}{\eta}(x,t)$. If $X(t)$ solves the (forward) SDE with drift \eqref{eq3.15}
\begin{equation}\label{eq5.4}
dX(t)=\hbar\nabla\log\eta(X(t),t)dt+\hbar^{1/2}dW_t
\end{equation}
and, therefore, the associated a.s Euler-Lagrange Eq \eqref{eq4.14},
\begin{equation}\label{eq5.5}
D_t D_t X(t)=0
\end{equation}
one checks that $h_\alpha(X(t),t)$ is a positive martingale:
\begin{equation}\label{eq5.6}
D_t h_\alpha (X(t), t)=0.
\end{equation}
Eq \eqref{eq5.3} corresponds to a Scaling transformation of the starting process $X(t)$, namely
\begin{equation}\label{eq5.7}
X^\alpha(t)=x_\alpha(X(t_\alpha),t(t_\alpha))=e^\alpha X(e^{-2\alpha} t)
\end{equation}
where $t(t_\alpha)$ denotes the inversion of the time parameter transformation in \eqref{eq5.2}.

The drift of $X^\alpha(t)$ results of a Doob's $h$ transform of $X(t)$ whose martingale is $h_\alpha$. Denoting by $B^\alpha$ and $B$ the associated drifts, we find \cite{Lescot08}, using the definition of $h_\alpha$,
\begin{equation}\label{eq5.8}
B^\alpha(x,t)=B(x,t)+\nabla\log h_\alpha(x,t)=\hbar\nabla\log\eta_\alpha(x,t).
\end{equation}
In other words, all diffusions $X(t)$ solving Eq \eqref{eq5.5} enjoy the scaling transformation symmetry, interpreted here dynamically. The standard Wiener $X(t)=W_t$, whose $\eta_\alpha(x,t)=1$, therefore $h_\alpha=1$ and $B_\alpha=0$, is only one of them: with $\varepsilon=e^{-2\alpha}$, we recover its usual scaling (or ``self-similarity") law $W^\varepsilon(t)=\varepsilon^{-1/2} W(\varepsilon t)$. This is useful, for instance, in the computation of first passage times of any diffusions solving Eq \eqref{eq5.5}. A large collection of parabolic equations (with first order and potential terms) is, in fact, equivalent to the above free heat equation, in terms of symmetries \cite{Lescot08}, so that our argument is more general (cf. also \cite{Olver86}).

Almost sure dynamical equations like \eqref{eq4.13} or \eqref{eq4.16}, together with our stochastic Noether Theorem provide new (geometrical) relations between familiar stochastic processes, impossible to anticipate without them. But, what may seem more surprising, they provide as well new informations about Quantum Mechanics in Hilbert space. For instance, the solutions $(T,Q,\phi)=(0,t,x)$ of the (one dimensional) determining equations \eqref{eq4.26} for $V=0$, applied to the standard Wiener $X(t)=W_t$, show that the family of Brownian martingales correspond to a family of (time dependent) constant observables in $L^2(\mathbb{R})$ for the free particle. Cf. \cite{Zamb09}. Although elementary, this observation had not been made before. Even better, a naive analytic continuation in time from the symmetries of the parabolic equation to the one of Schr\"odinger equation provides a Quantum Theorem of Noether richer that the one mentioned in Textbooks, even in elementary cases \cite{Alb06}.

The geometrical content of our stochastic deformation is worth an investigation in itself. Consider, for instance, the deformation of the classical method of Characteristics. One of the most elegant representation of the classical Hamilton-Jacobi (HJ) equation and its symmetries is due to E. Cartan and makes use of the following Ideal of differential forms \cite{Harrison71} (We denote here by $x_i$ what was $\omega_i$, in Eq \eqref{eq4.28}, to avoid confusions with differential forms): 
\begin{equation}\label{eq5.9}
I_{HJ}=
\begin{cases}
\omega=p^i dx_i-E\,d\tau+dS\equiv\omega_{PC}+dS\\
\Omega=dp^i dx_i-dE\,d\tau\\
\beta=\left(E-\frac{1}{2}|p|^2-V(x)\right) dx_i\,d\tau
\end{cases}
\end{equation}
on a 9-dimensional space of independent variables $(x_i,p^i,S,\tau,E)\quad i=1,2,3$.

Since $d\beta=(-dx_i+p^i dt)\,d\omega$, it belongs to the Ideal generated by $\omega$, $\Omega$ and $\beta$. According to Cartan, in these conditions, i.e when the Ideal is closed under exterior differentiation, the geometry of $HJ$ equation can be completely analyzed and this equation is ``integrable".

So \eqref{eq5.9} is Cartan's representation for the elementary systems treated after Eq \eqref{eq4.28}. Any solution of HJ will annul \eqref{eq5.9}. More precisely, to recover HJ equation itself, consider the $\mathbb{R}^4$ ``solution submanifold" where the a priori independent variable $S$ becomes a function $S(x,t)$ (This is called ``Sectioning" and denoted by $\sim$) and then pullback all differential forms to zero (``Annuling"). Then $\tilde\omega=0$ implies that $p=-\nabla S$ and $E=\partial_\tau S$. The condition $\tilde\Omega=0$ is equivalent to the existence of a Lagrangian manifold. Finally, $\tilde{\beta}=0$ is equivalent to the classical Hamilton-Jacobi itself. This representation shows that Hamilton-Jacobi framework is a Contact Geometry, defined on an odd dimensional space, here $\mathbb{R}^9$; $\omega$ is, in fact, a Contact form \cite{Lescot08}.

Cartan's theory of such ``Exterior differential systems" has been elaborated a lot in recent years.  A very thorough exposition can be found in \cite{BryGrifGros08}. One of the bonuses of a representation of HJ (more generally any pdes) as ideal of differential forms is the study of its symmetries, even those involving dependent and independent variables. In addition the result is, of course, coordinates invariant. \cite{Harrison71} provides many examples.

A symmetry generator of HJ equation become a priori a ``contact Hamiltonian" vector field (sometimes called ``Isovector")
\begin{equation}\label{eq5.10}
N=N^\tau\frac{\partial}{\partial\tau}+N^x_i\frac{\partial}{\partial x_i}+N^S\frac{\partial}{\partial S}+N^E\frac{\partial}{\partial E}+N^p_i\frac{\partial}{\partial P^i}
\end{equation}
whose coefficients must be chosen so that, denoting by $\mathcal{L}_N$ the Lie derivative, or variation, along $N$:
\begin{equation}\label{eq5.11}
\mathcal{L}_N(I_{HJ})\subseteq I_{HJ}.
\end{equation}
The stochastic deformation of $I_{HJ}$ is the one providing Hamilton-Jacobi Bellman equation \eqref{eq4.19}:
\begin{equation}\label{eq5.12}
I_{HJB}
\begin{cases}
\omega=P^i dX_i+E\,d\tau+dS\equiv \omega_{pc}+dS\\
\Omega=dP^i dX_i+dE\,d\tau\\
\beta=\left(E+\frac{1}{2}|P|^2-V\right) dX_i d\tau+\frac{\hbar}{2} dP^i d\tau.
\end{cases}
\end{equation}
The only deformation term, in $\beta$, is responsible for the deformation term $\frac{\hbar}{2}\Delta S$ in Eq \eqref{eq4.19}. Sectioning and annulling as before we find the Lagrangian integrability conditions: 
\begin{equation}\label{eq5.13}
\tilde\omega=0 \Rightarrow P=-\nabla S, E=-\partial_\tau S.
\end{equation}
The definition  of Symmetries for HJB is the same as classically, i.e Eq \eqref{eq5.11}, for the deformed ideal \eqref{eq5.12}, and the calculation of  the coefficients $N^\bullet$ of Eq \eqref{eq5.10} is a rather tiring exercise (cf. \cite{Lescot08}). But it is quite rewarding: 

\begin{thm}[\cite{Lescot08}]
Along any $N$-variation as before, $I_{HJB}$ and the Lagrangian $\mathcal{L}$ satisfy the following invariance conditions
\begin{align}\label{eq5.14}
\textrm{(1)} &\quad \mathcal{L}_N(\omega_{PC})=-dN^S\nonumber\\
\textrm{(2)} &\quad \mathcal{L}_N(\Omega)=0\\
\textrm{(3)} & \quad \mathcal{L}_N(\mathcal{L})+\mathcal{L}\frac{dN^\tau}{d\tau}=-D_\tau N^S.\nonumber
\end{align}
\end{thm}
This Theorem seems purely algebraic but encodes a lot of informations about our stochastic deformation, resulting from the substitution of smooth classical paths $\tau\mapsto \omega(\tau)$ by Bernstein diffusion sample paths $\tau\mapsto X(\tau)$. Eq (1) means that Poincar\'e-Cartan 1-form is invariant up to a phase coefficient $N^S$. Eq (2) shows the invariance of the Symplectic form over the time-dependent or extended phase space (cotangent bundle). Eq (3) expresses the transformation of the integrand of the Action functional \eqref{eq4.6} under the contact Hamiltonian $N$ on the extended phase space. It should be regarded as the deformation of the classical expression \eqref{eq4.27}.

The proof of the Theorem shows that it is, in fact, sufficient to consider symmetry contact Hamiltonians of the form $N(\tau,x,S,E,P)=N^x(x,\tau)P+N^\tau(x,\tau) E+N^S(x,\tau)$, so that $N^x=Q$, $N^\tau=T$ and $N^S=-\phi$ in the notations of the stochastic Noether Theorem, where $T,Q$ and $\phi$ solve its Determining Equations. After sectioning on the solution submanifold $(x,\tau)$ where, according to \eqref{eq3.9}, we have a probabilistic interpretation by plugging $x=X(\tau)$, $\tilde{P}$ and $\tilde{E}$ become respectively our drift and energy random variables.

The whole construction summarized before is preserved  if the diffusions $X(\tau)$ lives on a (smooth, connected, complete) $n$-dimensional Riemannian manifold with metric $g^{ij}$. The simplest Hamiltonian in Eq \eqref{eq4.24} becomes
\begin{equation}\label{eq5.15}
H=-\frac{\hbar^2}{2}\nabla^j\nabla_j + V(x)
\end{equation}
where $\nabla_j$ denotes the covariant derivative with respect to the L\'evi-Civita connection $\Gamma^i_{jk}$ and the associated Hamilton-Jacobi-Bellman equation \eqref{eq4.19} turns into
\begin{equation}\label{eq5.16}
\frac{\partial S}{\partial t}-\frac{1}{2}\|\nabla S\|^2 +\frac{\hbar}{2} \nabla^i\nabla_i S+V=0.
\end{equation}
Two new geometric aspects deserve to be mentioned. The first one is that (as already stressed by K. It\^o \cite{Ito63}) an additional term shows up in the drift \eqref{eq3.15}
\begin{equation}\label{eq5.17}
D_t X^i=\hbar\nabla^{i}\log\eta_t(X)-\frac{\hbar}{2}\Gamma^i_{jk}g^{jk}
\end{equation}
for $E_t dX^i dX^j=\hbar g^{ij} dt$.

On such a Riemannian manifold, an almost sure Euler-Lagrange equation like \eqref{eq4.14} (when $A=0$) requires to define the time derivative of a vector field. Even in the classical, deterministic case, a notion of parallel transport is needed to do that.According to It\^o \cite{Ito63}, the stochastic deformation of the L\'evi-Civita transport of the vector field $Y$ would transform Eq \eqref{eq4.12} into
\begin{equation}\label{eq5.18}
D_\tau Y^i=\frac{\partial Y^i}{\partial\tau}+\hbar\nabla^k\log\eta_t\nabla_k Y^i+\frac{\hbar}{2}\nabla^k\nabla_k Y^i.
\end{equation}
But It\^o also indicated other possible choices. The one needed for our purpose has been called ``Damped parallel transport" in Stochastic Analysis (cf. \cite{Mall97}), and replaces the Laplace-Beltrami term of \eqref{eq5.18} by 
\begin{equation}\label{eq5.19}
(\Delta Y)^i=\nabla^k\nabla_k Y^i+R^i_k Y^k
\end{equation}
where $R^j_k$ denotes the Ricci tensor. Then 
\begin{equation}\label{eq5.20}
D_\tau Y^i=\frac{\partial Y^i}{\partial\tau}+\hbar\nabla^k\log\eta_t\nabla_k Y^i+\frac{\hbar}{2}(\Delta Y)^i.
\end{equation}
The point is that to preserve for \eqref{eq5.16} the integrability condition according to which the gradient of \eqref{eq5.16} coincides with the Euler-Lagrange equation we need that $[\Delta,\nabla^i]S=0$, a property not satisfied by $\nabla^k\nabla_k$. Then, with \eqref{eq5.20}, the dynamical law and the Noether Theorem keep the same form as above \cite{Zambrini99}.

\section{Conclusions}
In 1985-6, I named after Bernstein the reciprocal property suggested by him in the context of the 1931 observation (forgotten until then) of Schr\"odinger \cite{Schr32}. I was, in fact, so impressed by the Bernstein interpretation of such processes as stochastic counterparts of critical trajectories of Hamilton's principle that I used as well the term ``variational processes" \cite{Zamb86}. Of course, the local Markov property reappeared during the seventies in relation  to Statistical and Quantum Physics. But Schr\"odinger's observation and Bernstein's probabilistic suggestion were, among other ideas, extraordinary anticipations of the Feynman Path Integral approach at the time.

There is more than one way to interpret Schr\"odinger's original observation, expressed originally in a statistical mechanics perspective, i.e in entropic terms. Aware of the fierce fights regarding the physical interpretation of the new born quantum theory, Schr\"odinger was looking for a ``classical" analogy where probabilities would play a similar but less debated role. In the seventies, B. Jamison \cite{Jam74} elaborated some aspects of the construction suggested  by Bernstein, but missed the sought relation with quantum theory (he was, for instance, using only the increasing $\mathcal{P}_t$ filtration and would not start his construction from a given Hamiltonian $H$). In any case, since the mid-eighties, Bernstein processes have reappeared in a multitude of contexts, pure and applied, and under different names. They were called ``Schr\"odinger processes" (following Jamison) by H. F\"ollmer \cite{Foll88} in 1988 and studied on their own in the entropic perspective \cite{Cruz00}, \cite{Leo10}, \cite{Walko89} and \cite{Daw90}. A promising link has been established with Optimal transport in recent years. An excellent review of this connection can be found in \cite{Leonard}. In this context, the natural approach is indeed the one of statistical mechanics, and the original variational problem in \cite{Schr32} is called ``Schr\"odinger 's problem" (not to be confused with Eq \eqref{eq3.18} referred to as the ``Schr\"odinger system"). ``Schr\"odinger bridges" is also a terminology used for these processes. They can really be regarded as a generalization of usual bridges where, instead of two boundary Dirac distributions, we are now given two arbitrary regular (nodeless) probability distributions. In recent studies of Wiener space, they have also proved to be quite natural tools \cite{Lassalle}.

Reciprocal Bernstein processes can also be characterized by an integration by parts formula, typical of Stochastic Analysis, but even when they are not Markovian \cite{Roelly}. It seems, indeed, that Feynman's symbolic approach was too limited to the Markovian class, not appropriate in many interesting cases. 

As it is clear from the first and last part of Section 4, we need three kinds of stochastic integrals for the complete description of Bernstein processes. A very general approach to stochastic integration with some similarities is due to Russo and Vallois cf. \cite{Russo93}. It would be interesting to reconsider our construction with the tools described therein.

The symmetry of such processes, in the sense of the Noether Theorem, can also be of interest for other purposes in probability theory \cite{Alili10}.

Clearly, the approach chosen here can be regarded as a random version of Geometric Mechanics (cf. \cite{LazOrt88}, \cite{Zamb09}, \cite{PrivZam10}). In this context, one of the most interesting open problems is the notion of Integrability suited to the random dynamical systems resulting from our approach. Some aspects of it have been used in the Ideal of differential forms \eqref{eq5.12}, but a lot more work remains to be done (cf. \cite{CruzWuZam}) in this direction.

It would also be interesting to understand the relations between our stochastic deformation and the (deterministic) deformation of characteristics for Hamiltonian PDEs inspired by B. Dubrovin \cite{Masoero}. One can, indeed, guess the existence of common features for some particular PDEs.

Some probabilists would question a study of such a special class of Bernstein reciprocal processes. The first reason is that this class is not as small as it seems. We hope that we have made clear that the key elements of their construction are independent of the form of the starting Hamiltonian $H$. Besides those like \eqref{eq3.25}, for L\'evy processes, we claim that it is always possible to time-symmetrize regular stochastic processes the way we have done.

The second reason is that it is precisely because this class is special that it carries all the qualitative properties needed to construct stochastic dynamical theories, which are generally missing in the regular approach to stochastic processes. As mentioned in $\S$2, Stochastic Analysis did not go at all in this direction. But, as suggested by the Feynman Path Integral approach, this direction seems to be the most natural one as far as physical theories are concerned.

There are many fields, outside Mathematics, where this unorthodox way to approach stochastic dynamics is often natural. For instance, in Finance \cite{Lescot}, Econometry \cite{Galichon} and Political Economy \cite{DupGal}. Image processing is also a promising domain of application for these ideas. A technical reason is obvious: the use of ``backward" heat equations like \eqref{eq3.11} (a very unfortunate terminology in our context) is common in image enhancement, although this PDE is ill posed in Hadamard's sense. This is due to the effect of deconvolution in the context of signal and image processing, where diffusion processes appear naturally. These diffusions must enjoy conflicting properties: they must simultaneously enhance, sharpen  and denoise images. Everything suggests that Bernstein processes are reliable and rigorous candidates for the job.

Of course, various problems of Statistical Physics can benefit as well from the use of time reversible probability measures \cite{Por11}. P.O. Kazinski, among many others, considers various classical models in this perspective \cite{Kaz08}. He also introduced the expression ``Stochastic Deformation". Other applications in Theoretical Physics include \cite{Garba95}, \cite{Duda}.

Random walks on graphs are described by Markov chains, reversible in a much narrower sense than the one intrinsic to Bernstein processes. It is likely that the methods used here will also be relevant in this area \cite{Radek09}. Interesting links with physics are explored in \cite{Duda}.

In Applied Mathematics, the relations of the variational component of the program summarized here and Stochastic Control Theory are, of course, striking. They strongly suggest that there are very few ways to deform systematically classical mechanics along diffusion processes. But these relations are still far from completely explored. It is remarkable, as mentioned above, that some investigations of the 1970-80's, aiming at a deformation of the classical calculus of variations along diffusion processes, were able to obtain results consistent with our probabilistic reinterpretation of Feynman's approach. What U.G. Haussmann \cite{Hauss81} calls the adjoint process, for instance, is basically our (forward) momentum process \eqref{eq4.7}. Of course, the results were all expressed with respect to a single (increasing) filtration, and as such were not directly appropriate to a time reversible dynamical framework.

A last comment about the Stochastic Deformation program: in the late 1960's, V. Arnold proved that the Euler equation of an ideal incompressible fluid could be interpreted in a (Lagrangian and Hamiltonian) analogy with the motion of a rigid body \cite{Arn69}. The configuration space was, then, the group of volume-preserving diffeomorphisms of the region occupied by the fluid. If we are not only interested in ``dry water" (as Feynman called the fluid described by Euler equation \cite{Feyn63}) then we have to deal with the Navier-Stokes equation. The idea that the equation corresponds to a stochastic deformation of the Euler equation was introduced informally in \cite{Naka81}, and has been made rigorous and considerably elaborated in recent years \cite{CruzArn}.

This means that the method of stochastic deformation can also be applied to some infinite dimensional and dissipative dynamical systems. Various books have already been published, where Bernstein reciprocal processes play a major role. We mention only two recent ones besides \cite{Chung03}: \cite{Guli06}, \cite{Cast11}. In each theoretical or experimental scientific situation where it seems natural to provide a pair of arbitrary initial and final probability densities, for a given system driven by any ``Hamiltonian" $H$, time reversible processes like those described here should arise.

This is why we are convinced that, also on the applied side, these processes do not only have a curious past  but also, indeed, a bright future.

\subsection*{Aknowledgment} 
It is a pleasure to thank the organizers of the stimulating semester program ``Stochastic Analysis and Applications" at EPFL, 9/1/2012 -- 30/6/2012, namely Robert Dalang, Marco Dozzi, Franco Flandoli and Francesco Russo. The active participation of the audience helped to improve the presentation of the material and is also warmly acknowledged.

\end{document}